\RequirePackage{silence}
\documentclass[english,twoside,openright]{UH_TCM_MSc}

\usepackage[T1]{fontenc}

\usepackage{amsfonts} 
\usepackage{amsmath, amssymb} 
\usepackage[titletoc]{appendix}
\usepackage{babel}
\usepackage[backend=biber,sorting=none]{biblatex}
\usepackage{blindtext}
\usepackage{bm} 
\usepackage[footnotesize,bf]{caption} 
\usepackage{csquotes} 
\usepackage[en-GB]{datetime2}
\usepackage{fancyhdr} 
\usepackage{color, graphicx} 
\usepackage[plainpages=false]{hyperref} 
\usepackage{listings}
\usepackage{lmodern} 
\usepackage{placeins} 
\usepackage{textcomp} 
\usepackage{tikz} 
\usepackage{titlesec}
\usepackage{todonotes}
\usepackage{xcolor}

\title{The effect of sound speed on the gravitational wave spectrum of first order phase transitions in the early universe}
\author{Mika Mäki}
\date{18th December 2024}
\newcommand{\dateupdated}{Updated on 25th November 2025}
\prof{Professor Mark Hindmarsh}
\censors{Professor Mark Hindmarsh}{Professor Kari Rummukainen}{}
\keywords{phase transitions, gravitational waves, sound speed, Sound Shell Model}
\depositeplace{arXiv, \href{https://hdl.handle.net/10138/591514}{Helda}, \href{https://github.com/AgenttiX/msc-thesis2}{GitHub}}
\additionalinformation{}

\classification{}

\hypersetup{
    unicode=true,           
    pdftoolbar=true,        
    pdfmenubar=true,        
    pdffitwindow=false,     
    pdfstartview={FitH},    
    pdftitle={The effect of sound speed on the gravitational wave spectrum of first order phase transitions in the early universe}, 
    pdfauthor={Mika Mäki},           
    pdfsubject={},          
    pdfcreator={},          
    pdfproducer={pdfLaTeX}, 
    pdfkeywords={phase transitions} {gravitational waves} {sound speed} {Sound Shell Model},
    pdfnewwindow=true,      
    colorlinks=true,        
    linkcolor=black,        
    citecolor=black,        
    filecolor=magenta,      
    urlcolor=cyan           
}

\begin{document}

\maketitle


\begin{abstract}
The Standard Model of particle physics has proven to be remarkably accurate in various collider experiments,
but lacks explanations for some observed phenomena such as baryon asymmetry: Why is there more matter than antimatter?
The extensions of the Standard Model provide possible solutions to these questions,
but the energy scales required to distinguish between them are difficult to achieve in colliders.
However, many of these extensions also result in cosmological phase transitions that have occurred early in the Big Bang during the electroweak symmetry breaking at around $10^{-11}$ s or earlier.

If these phase transitions have been of first order,
they have created a stochastic background of gravitational waves that can be directly observed today.
The Laser Interferometer Space Antenna (LISA) space probe will be launched in the 2030s to study gravitational waves and
to see whether such a stochastic background exists.
If we see a signal of cosmological origin,
we need to be able to deduce the parameters of the phase transition from the gravitational wave signal to understand the physics behind it.
To accomplish this, we need simulations of the gravitational wave spectra with various parameters.
One such important parameter is the sound speed $c_s$.

The vast majority of existing simulations have been based on the bag model equation of state,
which assumes the ultrarelativistic sound speed $c_s =\frac{1}{\sqrt{3}}$ for both phases.
This was also the case for the PTtools phase transition simulation framework developed by Hindmarsh et al.
In this thesis PTtools has been extended to include support for arbitrary equations of state and therefore for a temperature- and phase-dependent sound speed $c_s(T,\phi)$.
Since the sound speed is such an integral part of the hydrodynamic equations,
this required a nearly complete rewrite and significant extension of the code.
The code has also been sped up considerably by the use of the Numba JIT compiler, various other optimisations and parallelisation,
and made conformant to modern coding standards.

PTtools was tested with the constant sound speed model, in which the sound speed is a constant for each phase.
The sound speed was shown to have a significant effect on the resulting gravitational wave spectrum,
especially when changing the sound speed resulted in a change in the type of the solution.
This has laid the groundwork for simulating cosmological phase transitions with realistic equations of state based on the extensions of the Standard Model.
This will result in gravitational wave spectra that can be used in the LISA data analysis pipeline to search for the existence and parameters of a first-order phase transition in the early universe.

\end{abstract}

\mytableofcontents

\mynomenclature

\clearpage
\listoffigures
\clearpage
\listoftables
\clearpage



\chapter{Introduction}
\label{ch:introduction}
The cosmic microwave background (CMB) has been our primary source of information on the earliest moments of the universe.
It is the light from the time when electrons combined with nuclei to form atoms,
and the universe became transparent to electromagnetic radiation.
However, this occurred when the universe was about 370 000 years old,
and therefore so far we haven't been able to make direct observations from earlier events.
This is about to change in the 2030s, when the Laser Interferometer Space Antenna (LISA) will be launched.
It is a space probe that consists of three satellites arranged in an equilateral triangle with sides of 2.5 million kilometers,
and connected by laser interferometer beams.
LISA is designed to detect gravitational waves,
for which the universe has been transparent ever since gravity separated from the other three fundamental interactions.
Therefore, with gravitational waves we can make direct observations of events that occurred early in the Big Bang.
Possible sources of gravitational waves in the early universe include cosmological phase transitions, cosmic strings and inflation.
In this thesis I investigate the cosmological gravitational wave background generated by cosmological phase transitions.
\cites{lecture_notes}{lisa_2017}{colpi_lisa_2024}

In the Standard Model the Higgs transition is a cross-over instead of first-order.
Therefore, there is no sharp discontinuity at the phase boundary, and no generation of gravitational waves.
However, we know that the Standard Model is not a complete description of physics,
but has to be extended to account for various observations such as neutrino oscillations, matter-antimatter asymmetry and dark matter.
In many of these extensions of the Standard Model, the Higgs transition is a first-order phase transition
and therefore results in a gravitational wave signal that can be observed today.
This makes LISA a direct experiment that can distinguish between different extensions of the Standard Model.
The temperatures in the early universe were so extreme that they are difficult to replicate in particle colliders,
and therefore LISA is capable of probing conditions beyond the scope of existing particle colliders.
\cites{lecture_notes}{kajantie_is_1996}{caprini_detecting_2020}

First-order phase transitions proceed by the nucleation, expansion and collision of bubbles.
As the universe cools down,
it becomes energetically favourable for the field to change its phase.
However, in first-order phase transitions there is a potential barrier which prevents the field from transitioning to the new phase immediately.
Eventually this barrier is overcome by spontaneous quantum tunneling or thermal activation at random locations.
This releases energy, which pushes the nearby regions to the new phase as well.
This creates spherically expanding regions of the new phase surrounded by a phase boundary.
These are known as bubbles.
As the bubbles expand, they eventually collide and merge.
However, this is not the end of the story,
as the expanding domain walls have caused the fluid to move with them,
and these waves persist beyond the end of the phase transition.
These waves are strong enough to cause the space-time itself to ripple with them,
and these are the gravitational waves that, if they occurred, have persisted to the present day.
However, since they occurred in the very early universe, they have redshifted to much longer wavelengths in the millihertz regime.
The existing LIGO and Virgo gravitational wave detectors are ground-based detectors,
and their arms have been too short to detect these long-wavelength gravitational waves.
To detect the millihertz-range gravitational waves of cosmological origin, the much longer lasers of the LISA are required.
\cites{lecture_notes}{hindmarsh_gw_pt_2019}{mazumdar_review_2019}

The spectrum of the gravitational waves is characterised by five key parameters.
These are the nucleation temperature $T_n$,
phase transition strength at the nucleation temperature $\alpha_n$,
bubble wall speed $v_{\text{wall}}$,
transition rate parameter $\beta$
and the sound speed $c_s$.
\cite{lecture_notes}
The effects of the first four parameters have been studied extensively in the literature,
but in the vast majority of the studies so far with a few exceptions
\cites{leitao_hydrodynamics_2015}{giese_2020}{giese_2021}{tenkanen_speed_2022}{tian_gw_2024}
the sound speed $c_s$ has been assumed to be that of ultrarelativistic plasma: $c_s = \frac{1}{\sqrt{3}}$.
In this thesis I investigate more realistic scenarios, where the plasma is not fully ultrarelativistic,
but instead it has degrees of freedom $g(T,\phi)$ that are dependent on the temperature,
and depending on the definition of the model, also the phase,
and the potential of the field $V(T,\phi)$,
which depends on the temperature and the phase.
The speed of sound $c_s(T,\phi)$ is dependent on these quantities,
and therefore it's dependent on the temperature and the phase.
This complicates the numerical simulation of the fluid profile of the bubbles significantly.
\cites{leitao_hydrodynamics_2015}{giese_2020}{giese_2021}
This thesis is based on the phase transition simulation framework PTtools,
originally developed by Hindmarsh et al.
In this thesis it has been extended to account for these more complex models.
This updated version of PTtools and its documentation are available on GitHub \cite{pttools}.

This pdf, its LaTeX source code and the data analysis code of this thesis are available online on GitHub at \cite{thesis_source}.
Since this thesis also serves as an introduction and a part of the documentation of PTtools,
which continues to be developed,
this thesis may receive small updates after its publication.
Therefore, for the latest version, please see the GitHub repository \cite{thesis_source}.
This thesis is licensed with
\href{https://creativecommons.org/licenses/by/4.0/}{Creative Commons Attribution 4.0 International}.

\chapter{Phase transitions in the early universe}
\label{ch:pt}
The early universe has undergone multiple phase transitions.
Among the most notable of these is the electroweak symmetry breaking, also known as the Higgs transition, at around $10^{-11}$ s and $100$ -- $1000$ GeV,
when the electromagnetic interaction separated from the weak interaction and the Higgs mechanism turned on,
giving the gauge bosons their rest masses.
Another notable phase transition is the QCD phase transition at around $10^{-5}$ s and $200$ MeV,
when the quarks of the quark-gluon plasma of the early universe became bound into hadrons.
In the Standard Model these are both second-order phase transitions, also known as crossovers.
\cites{lecture_notes}{aoki_order_2006}

The electroweak symmetry breaking is of particular interest,
as the transition is of first order in various extensions of the Standard Model.
A first-order phase transition results in a departure from thermal equilibrium at the phase boundary,
which is a requirement for the formation of matter-antimatter asymmetry.
This generation of net baryon number is known as baryogenesis.
\cite{lecture_notes}

The theory of first-order phase transitions in the early universe is based on relativistic hydrodynamics.
The following section~\ref{rel_hydro} starts from the energy-momentum tensor of general relativity and derives the equations that
govern the evolution of the fluid: the relativistic Euler equation, the energy-conservation equation and the wave equation.
Section~\ref{relativistic_combustion} proceeds to the process of relativistic combustion and derives the bubble wall junction conditions
that define the behaviour of the fluid velocities and thermodynamic quantities at the bubble wall.
It then defines various equations and quantities that will be of use in the simulations,
including the transition strength $\alpha$ and an ordinary differential equation group that determines the fluid velocity profile of the bubble.
Section~\ref{solution_types} investigates the different types of solutions that these equations allow,
and the speed limits associated with these solutions.
Section~\ref{general_eq} derives the general equation of state,
and section~\ref{bag_model} introduces its simplest implementation, the bag model.
Section~\ref{const_cs} extends the bag model to having different sound speeds in each phase,
resulting in the constant sound speed model.
Finally section~\ref{energy_redistribution} investigates how the phase transition affects the distribution of energy between the field and the kinetic and thermal energies of the fluid.

\section{Relativistic hydrodynamics}
\label{rel_hydro}

Our system of interest is an ultrarelativistic plasma.
This means, that the energy of the particles is much higher than their rest mass.
Therefore, there is sufficient energy for new particles to be created,
and similarly existing particles will annihilate all the time.
Treating this kind of a system as a classical fluid is not sufficient,
and we need the mathematical machinery of general relativity.
It should be noted that the Einstein notation will be used for the indices throughout the thesis.
Greek indices are used for four-vectors and latin indices for three-vectors.

In general relativity the matter and energy content of space are described by the energy-momentum tensor $T_{\mu \nu}$, also known as the stress-energy tensor.
In Minkowski space in Cartesian coordinates this can be expressed nicely as a matrix,
if we take the time direction to be the coordinate time, $u^\mu = \delta^{\mu 0}$.
Then it's given as
\cites[eq. 4.17]{rasanen_gr_2022}[fig. 3.3]{rezzolla_relativistic_2013}
\begin{equation}
T_{\mu \nu} =
\begin{bmatrix}
e & -q_1 & -q_2 & -q_3 \\
-q_1 & p + \Pi_{11} & \Pi_{12} & \Pi_{13} \\
-q_2 & \Pi_{12} & p + \Pi_{22} & \Pi_{23} \\
-q_3 & \Pi_{13} & \Pi_{23} & p + \Pi_{33}
\end{bmatrix},
\label{eq:ep_tensor_general_matrix}
\end{equation}
where $e$ is the energy density, $p$ is the isotropic pressure, $q$ is the energy flux or momentum density and
$\Pi_{ij}$ is known as the anisotropic stress, anisotropic pressure or momentum flux,
for which $\delta^{ij} \Pi_{ij} = 0$.
Therefore the pressure can be extracted as
\begin{equation}
p = \frac{1}{3} T^i_i.
\label{eq:pressure_from_ep_tensor}
\end{equation}

In our case we assume the energy-matter content to be an ideal fluid in thermal equilibrium.
There is no energy transfer to or from the fluid, and therefore
$\forall j=1,2,3: \ T^{0j} = 0$.
As the fluid has no viscosity, it does not experience any shear stress, and therefore
$\forall i,j=1,2,3, \ i \neq j: \ T^{ij} = 0$.
This does not depend on the reference frame, so the tensor is diagonal in the rest frame of the fluid.
Therefore $T^{ij} = p \delta^{ij}$, which means that
$\forall i,j: \Pi_{ij} = 0$.
These are satisfied only by the tensor $T^{00}=e, T^{jj}=p, \forall i \neq j: T^{ij}=0$, which in matrix form is
\begin{equation}
T_{\mu \nu} =
\begin{bmatrix}
e & 0 & 0 & 0 \\
0 & p & 0 & 0 \\
0 & 0 & p & 0 \\
0 & 0 & 0 & p
\end{bmatrix}.
\end{equation}
This
energy-momentum tensor of an ideal fluid can be broken in two components as
\cites[eq. 5.11, 5.23]{lecture_notes}[eq. 4.12]{rasanen_gr_2022}
\begin{align}
T^{\mu \nu}_f
&= (e+p) u^\mu u^\nu + p g^{\mu \nu}
\label{eq:ep_tensor} \\
&= w u^\mu u^\nu + p g^{\mu \nu}.
\end{align}
This expression is independent of our choice of a coordinate system.
We have also defined the enthalpy density $w = e+p$ to simplify the expression.
The enthalpy density will be defined more thoroughly later in~\eqref{eq:enthalpy_sum}.

We are assuming the background space-time to be constant, and therefore the total energy-momentum is conserved.
In the language of general relativity this is
\begin{equation}
\nabla_\mu T^{\mu\nu} = 0.
\label{eq:ep_conservation}
\end{equation}
Here $\nabla_\mu$ is the covariant derivative, which in Minkowki space in Cartesian coordinates reduces to the partial derivative $\partial_\mu$.
To expand this equation in a way consistent with~\cite[ch. 3.3]{rezzolla_relativistic_2013} we need to define the projection tensor
\cite[eq. 3.9]{rezzolla_relativistic_2013}
\begin{equation}
h_{\mu\nu} \equiv g_{\mu\nu} + u_\mu u_\nu,
\label{eq:projection_tensor}
\end{equation}
and the expansion scalar
\cite[eq. 3.13]{rezzolla_relativistic_2013}
\begin{equation}
\Theta \equiv h^{\mu\nu} \nabla_\nu u_\mu = \nabla_\mu u^\mu.
\end{equation}
Inserting these to~\eqref{eq:ep_conservation} results in
\begin{equation}
\nabla_\mu T^{\mu \lambda} = u^\lambda u^\mu \nabla_\mu w + w u^\mu \nabla_\mu u^\lambda + w \Theta u^\lambda + g^{\lambda\mu} \nabla_\mu p.
\end{equation}
Relativistic acceleration is defined as
\begin{equation}
a_\nu = u^\mu \nabla_\mu u_\nu.
\end{equation}
It is orthogonal to the velocity, and therefore
\begin{equation}
a^\mu u_\mu = 0.
\end{equation}
Projecting using~\eqref{eq:ep_conservation} and using these we get
\cite[eq. 3.54]{rezzolla_relativistic_2013}
\begin{equation}
\bm{h} \cdot \bm{\nabla} \cdot \bm{T}
= h^\nu_\lambda \nabla_\mu T^{\mu \lambda}
= w u^\mu \nabla_\mu u^\nu + h^\nu_\lambda g^{\lambda\mu} \nabla_\mu p
= 0.
\end{equation}
It should be noted that~\cite{rezzolla_relativistic_2013} uses the specific enthalpy $h$ in their form of these equations,
but it is not a meaningful quantity for an ultrarelativistic plasma and is therefore not used here.
Dividing by $w$ we get the relativistic Euler equations
\cite[eq. 3.55]{rezzolla_relativistic_2013}
\begin{equation}
u^\mu \nabla_\mu u_\nu + \frac{1}{w} h^\mu_\nu \nabla_\mu p = 0.
\end{equation}
Similarly, we can project along the direction of velocity,
\begin{equation}
\bm{u} \cdot \bm{\nabla} \cdot \bm{T} = u_\lambda \nabla_\mu T^{\mu\lambda} = 0.
\end{equation}
This results in the energy-conservation equation
\cite[eq. 3.57]{rezzolla_relativistic_2013}
\begin{equation}
u^\mu \nabla_\mu e + w \Theta = 0.
\end{equation}

For a one-dimensional flow in Cartesian coordinates the energy-momentum conservation of~\eqref{eq:ep_conservation} can also be rewritten as
\begin{align}
\partial_t \left[ (e+pv^2) \gamma^2 \right] + \partial_x \left[ (e+p) \gamma^2 v \right] &= 0,
\label{eq:ep_conservation_1d_1} \\
\partial_t \left[ (e+p) \gamma^2 v \right] + \partial_x \left[ (ev^2 + p) \gamma^2 \right] &= 0
\label{eq:ep_conservation_1d_2}
\end{align}
by simply inserting~\eqref{eq:ep_tensor} and using the normalisation of four-velocity $u_\mu u^\mu = -1$.
Let us then analyse a perturbation on a fluid that is at rest with $e_0, p_0$ and $v_0 = 0$.
To first order
\begin{equation}
e = e_0 + \delta e, \quad p=p_0 + \delta p, \quad v = v_0 + \delta v = \delta v.
\end{equation}
Substituting these into~\eqref{eq:ep_conservation_1d_1} and~\eqref{eq:ep_conservation_1d_2} and approximating to first order in the perturbation results in
\begin{align}
\partial_t (\delta e) + (e_0 + p_0) \partial_x (\delta v) = 0,
\label{eq:ep_conservation_rewritten_1} \\
\partial_t (\delta v) + \frac{1}{e_0 + p_0} \partial_x (\delta p) = 0.
\label{eq:ep_conservation_rewritten_2}
\end{align}
Taking the time derivative of~\eqref{eq:ep_conservation_rewritten_1} and a space derivative of~\eqref{eq:ep_conservation_rewritten_2}, we can combine the equations to the form
\begin{equation}
\partial_t^2 (\delta e) - \frac{\delta p}{\delta e} \partial_x^2(\delta e) = 0.
\label{eq:wave_equation}
\end{equation}
This is the familiar wave equation,
so the motion of the fluid has wave solutions that we call sound.
\cites[ch. 4.3]{rezzolla_relativistic_2013}[ch. 7.4]{lecture_notes}
The speed of sound is then defined as
\cites[eq. 2.168]{rezzolla_relativistic_2013}[eq. 13]{giese_2020}[eq. 3]{giese_2021}
\begin{align}
c_s^2
&\equiv \frac{dp}{de}
\label{eq:cs2_compact} \\
&= \frac{dp/dT}{de/dT}.
\label{eq:cs2_explicit}
\end{align}
At the ultrarelativistic limit $e=3p$ and therefore $c_s^2=\frac{1}{3}$, as we will derive in section~\ref{general_eq}.

Proper discussion of relativistic fluids also requires various thermodynamic quantities.
Perturbing around a constant internal energy $U$,
effectively setting $dU=0$,
gives the entropy density
\cite[p. 23]{lecture_notes}
\begin{equation}
s \equiv \frac{dS}{dV} = \frac{dp}{dT},
\label{eq:entropy_density}
\end{equation}
where $S$ is entropy, $V$ is volume and $T$ is temperature.
Useful formulas for other thermodynamic quantities can be obtained from the thermodynamic identity of classical thermal physics.
\cites[eq. 2.136]{rezzolla_relativistic_2013}[eq. 3.68]{schroeder_thermal_2000}
\begin{equation}
dU = TdS - pdV, 
\label{eq:thermodynamic_identity}
\end{equation}
The fluids in our case are ultrarelativistic,
and therefore effectively the temperature is much higher than the chemical potential $\mu$,
$T \gg \mu$,
which means that we can neglect $dN$.
Inserting~\eqref{eq:entropy_density} to~\eqref{eq:thermodynamic_identity} gives the energy density
\begin{equation}
e \equiv \frac{dU}{dV} = T \frac{\partial p}{\partial T} - p.
\label{eq:energy_density}
\end{equation}
This in turn can be inserted to the definition of enthalpy
\cite[eq. 1.51]{schroeder_thermal_2000}
\begin{equation}
H \equiv E + pV
\end{equation}
to get the enthalpy density
\begin{align}
w
&\equiv \frac{dH}{dV} \\
&= e+p
\label{eq:enthalpy_sum} \\
&= T \frac{\partial p}{\partial T}
\label{eq:enthalpy_pressure}\\
&= Ts.
\label{eq:enthalpy_entropy}
\end{align}
Additionally we need to define the entropy current
\cite[p. 23]{lecture_notes}
\begin{equation}
S^\mu \equiv su^\mu.
\end{equation}
These thermodynamic quantities are actively used in the following sections.

\section{Field-fluid system}
Our system of interest consists of a relativistic fluid of particles and the scalar field $\phi$,
which are coupled to each other.
The fluid is governed by the energy-momentum tensor of eq.~\eqref{eq:ep_tensor},
and the scalar field is governed by the energy momentum tensor
\cites[eq. 5.12]{lecture_notes}[eq. 2.9]{hindmarsh_gw_pt_2019}
\begin{equation}
T_\phi^{\mu \nu}
= \partial^\mu \partial^\nu \phi
- g^{\mu \nu} \left(\frac{1}{2} (\partial \phi)^2 + V_0 (\phi) \right).
\label{eq:ep_tensor_field}
\end{equation}
The total energy-momentum is conserved, and eq.~\eqref{eq:ep_conservation} can therefore be decomposed as
\cite[eq. 5.17]{lecture_notes}
\begin{equation}
\partial_\mu T^{\mu \nu} = \partial_\mu (T_f^{\mu \nu} + T_\phi^{\mu \nu}) = 0.
\label{eq:ep_conservation_total}
\end{equation}
For further details on how the field-fluid system dynamics arise from the underlying particle physics,
please see~\cites{lecture_notes}{moore_pt_1995}.

\begin{figure}[ht!]
\centering
\includegraphics[width=0.6\textwidth]{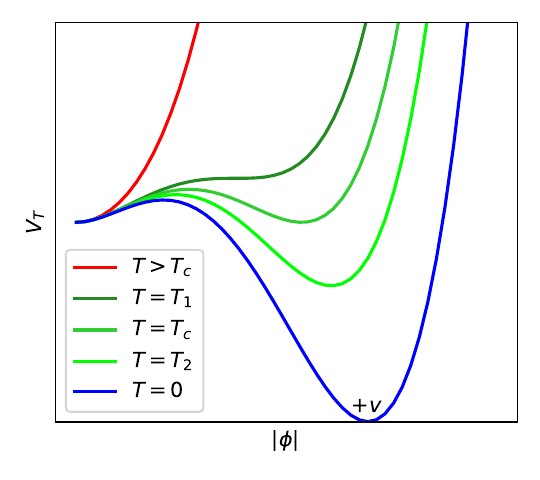}
\caption{The thermal effective potential at different temperatures. Generated with PTtools utilities. See also~\cite[fig. 4]{lecture_notes}.}
\label{fig:potential}
\end{figure}

Figure~\ref{fig:potential} illustrates the temperature-dependent shape of the thermally corrected potential $V_T$.
For temperatures $T > T_c$ (red) the minimum of the potential is at $\phi = 0$, and the ground state is symmetric.
As the temperature decreases below $T_1$ (dark green), a second minimum develops, but its potential is higher than that of the ground state.
At the critical temperature $T_c$ both minima have the same potential and are therefore degenerate.
Below the critical temperature, the minimum at $\phi > 0$ becomes the global minimum and therefore the new stable ground state.
This results in eventual overcoming of the potential barrier by quantum tunneling or thermal activation from the $\phi = 0$ state to the $|\phi| = v$ state,
and therefore the spontaneous nucleation of bubbles.

The pressure is dependent on the field,
since the field gives the particles their masses.
For the bubble to expand, the interior pressure has to be higher than the exterior one.
At the critical temperature $T_c$ of a phase transition,
the pressures on both sides of the wall are equal
\begin{equation}
p_+(T_c) = p_-(T_c),
\label{eq:critical_temp}
\end{equation}
where $+$ refers to the symmetric phase in front of the bubble wall,
and the $-$ refers to the broken phase behind the bubble wall.
For a more comprehensive description on the pressure, please see the section~\ref{general_eq}.
The nucleation temperature $T_n$ has to be below the critical temperature for the phase transition to occur: $T_n < T_c$.
The nucleation enthalpy $w_n$ is related to the nucleation temperature $T_n$ as
\begin{equation}
w_n = w(T_n,\phi_s).
\label{eq:wn}
\end{equation}
Below the critical temperature, the bubble expands to the surrounding plasma.
The dynamics of the field $\phi$ driving the phase transition are given by
\cite[eq. 7.2]{lecture_notes}
\begin{equation}
\square \phi - V'_T(\phi) = - \eta_T(\phi) u^\mu \partial_\mu \phi,
\end{equation}
where $\eta$ determines the strength of the friction due to the interaction of the bubble with the plasma.
The pressure difference between the symmetric and broken phases drives the expansion of the bubble,
and would lead to continued acceleration if not counteracted by the friction $\eta_T$.
In some cases the friction is not sufficient to slow the expansion down,
resulting in $v_\text{wall} \rightarrow 1$,
which is known as a runaway solution~\cites{bodeker_can_2009}{bodeker_electroweak_2017}.
However, if $\gamma_\text{wall} \eta_T$ does not decrease with $\gamma_\text{wall}$,
then there always exists a solution with a constant $v_\text{wall}$~\cites{bodeker_electroweak_2017}{hoche_towards_2021}.
The wall speed is further investigated in section~\ref{solution_types}.

\section{Relativistic combustion}
\label{relativistic_combustion}
Cosmological phase transitions are analogous to classical combustion in the sense
that both have an advancing transition front,
where energy is released,
and hydrodynamic waves that are generated by the transition front.
Therefore, they can be said to be a form of relativistic combustion.
This section investigates the behaviour of the field-fluid system at the phase boundary,
also known as the bubble wall,
and the evolution of the fluid in front and behind of the wall.
\cites[appendix B]{hindmarsh_gw_pt_2019}{espinosa_energy_2010}{rezzolla_relativistic_2013}[ch. 6]{mazumdar_review_2019}.

If the wall velocity is constant,~\eqref{eq:ep_conservation_total} also holds in the wall frame.
Therefore, for a wall moving in the z-direction
\cite[eq. 7]{espinosa_energy_2010}
\begin{equation}
\partial_z T^{zz} = \partial_z T^{z0} = 0.
\end{equation}
Inserting the definition of $T$ from~\eqref{eq:ep_tensor} results in the
bubble wall junction conditions
\cites[eq. 7.22]{lecture_notes}[eq. 135.2, 135.3]{landau_fluid_1987}
\begin{align}
w_- \tilde{\gamma}_-^2 \tilde{v}_- &= w_+ \tilde{\gamma}_+^2 \tilde{v}_+,
\label{eq:junction_condition_1} \\
w_- \tilde{\gamma}_-^2 \tilde{v}_-^2 + p_- &= w_+ \tilde{\gamma}_+^2 \tilde{v}_+^2 + p_+.
\label{eq:junction_condition_2}
\end{align}
The first junction condition can be trivially rearranged to
\begin{equation}
w_- = \frac{\tilde{\gamma}_+^2 \tilde{v}_+}{\tilde{\gamma}_-^2 \tilde{v}_-} w_+,
\label{eq:wm_junction}
\end{equation}
which we will need later.
Let us now define two intermediate quantities,
\begin{equation}
x \equiv \tilde{v}_+ \tilde{v}_-, \quad y \equiv \frac{\tilde{v}_+}{\tilde{v}_-}.
\end{equation}
Using these we can rearrange the first junction condition of eq.~\eqref{eq:junction_condition_1} as
\begin{equation}
\frac{w_-}{w_+} = \frac{y-x}{1-xy},
\label{eq:junction_step1}
\end{equation}
and the second junction condition of eq.~\eqref{eq:junction_condition_2} as
\begin{equation}
\frac{\Delta p}{w_+} (y-x) - \frac{w_-}{w_+} x + \frac{y-x}{1-xy} + (x-y) = 0.
\label{eq:junction_step2}
\end{equation}
Inserting eq.~\eqref{eq:junction_step1} to~\eqref{eq:junction_step2} we get
\begin{equation}
y = \frac{x \Delta e + w_-}{e_+ + p_-}.
\end{equation}
Inserting this back to eq.~\eqref{eq:junction_step1} results in the quadratic equation
\begin{equation}
(\Delta e) x^2 - (\Delta w) x + \Delta p = 0.
\end{equation}
Solving this and discarding the unphysical solution $x = 1$ results in
\cites[eq. 7.32]{lecture_notes}[eq. 4.134]{rezzolla_relativistic_2013}
\begin{align}
x &= \tilde{v}_+ \tilde{v}_- = \frac{p_+ - p_-}{e_+ - e_-}, \\
y &= \frac{\tilde{v}_+}{\tilde{v}_-} = \frac{e_- + p_+}{e_+ + p_-}.
\label{eq:junction_ep}
\end{align}
Further simplification of these requires additional definitions
that we will now introduce.


The trace of the energy-momentum tensor of~\eqref{eq:ep_tensor} is
\begin{align}
T^\mu_\mu
&= g_{\mu \nu} T^{\mu \nu} \\
&=3p - e.
\end{align}
We can quantify its difference from zero with the trace anomaly
\cites[eq. 7.24]{lecture_notes}[eq. 28]{giese_2020}
\begin{equation}
\theta \equiv \frac{1}{4}(e-3p).
\label{eq:theta}
\end{equation}
The definition of the trace anomaly varies by a factor of $\frac{1}{4}$ depending on the source as a matter of convention.
To alleviate this confusion, we introduce the trace anomaly without the factor of $\frac{1}{4}$ as
\begin{equation}
\Theta \equiv e - 3p.
\label{eq:theta_big}
\end{equation}
The difference of a quantity $X$ just ahead and behind the wall is denoted as
\begin{equation}
\Delta X
\equiv X_+(T_+) - X_-(T_-). 
\end{equation}
The difference of a quantity $X$ in both phases at the same temperature is denoted as
\begin{equation}
DX_\pm \equiv X_+(T_\pm) - X_-(T_\pm),
\end{equation}
where $\pm$ means that we can compute this quantity at either $T_+$ or $T_-$.
The difference of a quantity $X$ in the same phase at different temperatures is denoted as
\begin{equation}
\delta X_\pm \equiv X_\pm(T_+) - X_\pm(T_-).
\end{equation}
These are related by
\begin{align}
\Delta X
&= X_+(T_+) - X_-(T_-) \\
&= \left( X_+(T_+) - X_-(T_+) \right) + \left( X_-(T_+) - X_-(T_-) \right) \\
&= DX_+ + \delta X_-.
\label{eq:delta_relation}
\end{align}
With these we can define the transition strength
\begin{equation}
\alpha_+
\equiv \frac{4 \Delta \theta}{3 w_+}
= \frac{4}{3} \frac{\theta_+(T_+) - \theta_-(T_-)}{w_+}
\label{eq:alpha_plus}
\end{equation}
and the transition strength at nucleation temperature
\cite[eq. 2.11]{hindmarsh_gw_pt_2019}
\begin{equation}
\alpha_n
\equiv \frac{4D\theta(T_n)}{3w_n}
= \frac{4}{3} \frac{\theta_+(T_n) - \theta_-(T_n)}{w_n}.
\label{eq:alpha_n}
\end{equation}
A quantity similar to the trace anomaly is the pseudotrace
\cites[eq. 34]{giese_2020}[eq. 1]{giese_2021}
\begin{equation}
\bar{\Theta} \equiv e - \frac{p}{c_{s,b}^2}.
\label{eq:theta_bar}
\end{equation}
It is defined with respect to $c_{s,b}^2$ in both phases so that eq.~\eqref{eq:junction_conditions_simplified} can be simplified.
It should be noted that the articles \cites{giese_2020}{giese_2021} denote this pseudotrace as $\bar{\theta}$,
but we have chosen the symbol $\bar{\Theta}$ in accordance with~\eqref{eq:theta_big} to highlight that it doesn't have the factor of $\frac{1}{4}$ as eq.~\eqref{eq:theta} does.
Using it we can define the transition strength for the pseudotrace,
sometimes denoted as $\alpha_{\bar{\Theta}}$, as
\cites[eq. 34]{giese_2020}[eq. 1]{giese_2021}
\begin{equation}
\alpha_{\bar{\Theta}_+}
\equiv \frac{D \bar{\Theta}(T_+)}{3w_+}
= \frac{\bar{\Theta}_+(T_+) - \bar{\Theta}_-(T_+)}{3w_+}.
\label{eq:alpha_theta_bar_plus}
\end{equation}
The transition strength for the pseudotrace at the nucleation temperature is given as
\begin{equation}
\alpha_{\bar{\Theta}_n}
\equiv \frac{D \bar{\Theta}(T_n)}{3w_n}
= \frac{\bar{\Theta}_+(T_n) - \bar{\Theta}_-(T_n)}{3w_n}.
\label{eq:alpha_theta_bar_n}
\end{equation}
Note that these equations differ from the regular transition strength by a factor of $\frac{1}{4}$ due to different conventions,
but this is cancelled by the lack of $\frac{1}{4}$ in $\bar{\Theta}$.
These also use $D$ instead of $\Delta$,
and will prove out to be convenient with the constant sound speed model of section \ref{const_cs}.
We can also define the enthalpy ratio
\begin{equation}
r \equiv \frac{w_+}{w_-}.
\end{equation}
With these definitions the junction conditions of~\eqref{eq:junction_ep} can be simplified as
\begin{equation}
\tilde{v}_+ \tilde{v}_- = \frac{1-(1-3\alpha_+)r}{3-3(1+\alpha_+)r},
\quad
\frac{\tilde{v}_+}{\tilde{v}_-} = \frac{3+(1-3\alpha_+)r}{1+3(1+\alpha_+)r}.
\label{eq:junction_conditions_simplified}
\end{equation}
This is easiest to prove in the reverse direction.
To further get expressions for $\tilde{v}_+$ and $\tilde{v}_-$ that are independent of $r$, we can extract $r$ from both equations and mark those as equal, resulting in the equation
\begin{equation}
\left( \frac{1}{\tilde{v}_-} + 3 \tilde{v}_- \right) \tilde{v}_+ - 3(1+\alpha_+)\tilde{v}_+^2 = 1 - 3\alpha_+.
\end{equation}
This is of second order in both $\tilde{v}_+$ and $\tilde{v}_-$ and can be solved as \cite[eq. B.6, B.7]{hindmarsh_gw_pt_2019}
\begin{align}
\tilde{v}_+ &= \frac{1}{2(1+\alpha_+)}\left[ \left(\frac{1}{3\tilde{v}_-}+\tilde{v}_-\right) \pm \sqrt{\left(\frac{1}{3\tilde{v}_-} - \tilde{v}_- \right)^2 + 4\alpha_+^2 + \frac{8}{3} \alpha_+} \right],
\label{eq:v_tilde_plus}
\\
\tilde{v}_- &= \frac{1}{2} \left[ \left( (1+\alpha_+)\tilde{v}_+ + \frac{1-3\alpha_+}{3\tilde{v}_+} \right) \pm \sqrt{\left((1+\alpha_+)\tilde{v}_+ + \frac{1-3\alpha_+}{3\tilde{v}_+} \right)^2 - \frac{4}{3}} \right].
\label{eq:v_tilde_minus}
\end{align}
Both of these are basic second-order solutions, but in eq.~\eqref{eq:v_tilde_plus} the discriminant has been simplified.

It should be noted that the equations~\eqref{eq:junction_condition_1} and~\eqref{eq:junction_condition_2} are symmetric with respect to the indices $+$ and $-$.
We can define another transition strength by inverting the indices of $\alpha_+$ in~\eqref{eq:alpha_plus} as
\begin{equation}
\alpha_- \equiv - \frac{4 \Delta \theta}{3 w_-}.
\end{equation}
With this we can solve the junction conditions as above, but with opposite indices, resulting in
\begin{align}
\tilde{v}_+ &= \frac{1}{2} \left[ \left( (1+\alpha_-)\tilde{v}_- + \frac{1-3\alpha_-}{3\tilde{v}_-} \right) \pm \sqrt{\left((1+\alpha_-)\tilde{v}_- + \frac{1-3\alpha_-}{3\tilde{v}_-} \right)^2 - \frac{4}{3}} \right],
\label{eq:v_tilde_plus_reverse}
\\
\tilde{v}_- &= \frac{1}{2(1+\alpha_-)}\left[ \left(\frac{1}{3\tilde{v}_+}+\tilde{v}_+\right) \pm \sqrt{\left(\frac{1}{3\tilde{v}_+} - \tilde{v}_+ \right)^2 + 4\alpha_-^2 + \frac{8}{3} \alpha_-} \right].
\label{eq:v_tilde_minus_reverse}
\end{align}
These versions of the equations are useful when computing $\tilde{v}_+(\alpha_+)$ in a model where $\alpha_-$ is independent of $w_+$, as then $\tilde{v}_+ = \tilde{v}_+(w_-)$.

\begin{figure}[ht!]
\centering
\includegraphics[width=0.7\textwidth]{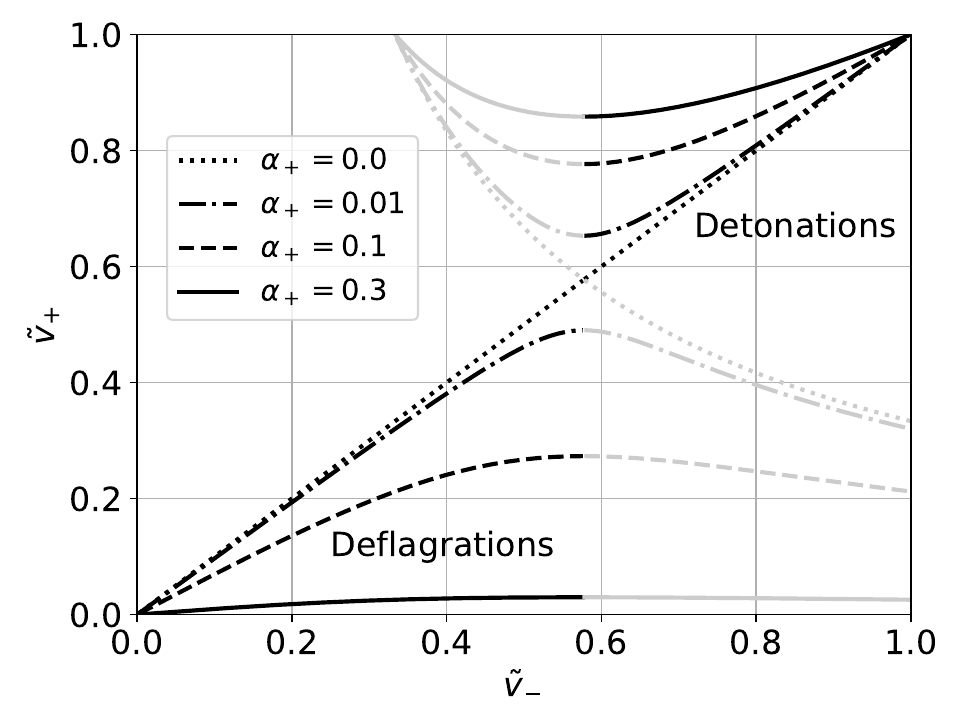}
\caption{Relation between the fluid velocity just ahead of the wall $\tilde{v}_+$ and just behind the wall $\tilde{v}_-$. Generated with PTtools. See also \cite[fig. 13]{lecture_notes}.}
\label{fig:vplus_vminus}
\end{figure}

We can obtain useful continuity equations from the conservation of energy-momentum.
We start by defining two vectors, the fluid 4-velocity
\begin{equation}
u^\mu \equiv \gamma(1, \overrightarrow{v})^\mu
\label{eq:u_mu}
\end{equation}
and its orthonormal vector
\begin{equation}
\bar{u}^\mu \equiv \gamma(v, \frac{\overrightarrow{v}}{v})^\mu.
\end{equation}
Please note that the way these vectors are defined and the convention for the signature of the metric vary
from article to article.
These vectors form an orthonormal basis.
It should be noted, that for orthonormal vectors
\cite[p. 36]{lecture_notes}
\begin{align}
u_\mu u^\mu &= 1, \\
u_\mu \bar{u}^\mu &= 0.
\end{align}
By projecting the conservation equation~\eqref{eq:ep_conservation} to this basis and noting that
$\nabla_\nu (u_\mu u^\mu) = 0$ and therefore $u_\mu \nabla_\nu u^\mu = 0$, we get
\cite[eq. 7.28-7.29]{lecture_notes}
\begin{align}
0 &= u_\mu \nabla_\nu T^{\mu \nu} = -\nabla_\mu (w u^\mu) + u^\mu \nabla_\mu p,
\label{eq:continuity1} \\
0 &= \bar{u}_\mu \nabla_\nu T^{\mu \nu} = w \bar{u}^\nu u^\mu \nabla_\mu u_\nu + \bar{u}^\mu \nabla_\mu p.
\label{eq:continuity2}
\end{align}
These are known as the continuity equations, or the hydrodynamic equations.
It should be noted that many articles use the symbol of a partial derivative $\partial$ in these equations,
even though it's actually a covariant derivative $\nabla$.
Since we are operating in spherical coordinates,
the divergence of a vector $A$ is given by
\begin{align}
\nabla \cdot A
&= \frac{1}{r^2} \frac{\partial (r^2 A^r)}{\partial r}
+ \frac{1}{r \sin \theta} \frac{\partial A^\theta}{\partial \theta}
+ \frac{1}{r \sin \theta} \frac{\partial A^\phi}{\partial \phi}.
\label{eq:divergence}
\end{align}
Let us look for a spherically symmetric self-similar solution using the dimensionless coordinate
\begin{equation}
\xi \equiv \frac{r}{t}.
\label{eq:xi}
\end{equation}
In a spherically symmetric case for $u^\mu$ of eq.~\eqref{eq:u_mu},
the latter two terms of eq.~\eqref{eq:divergence} are zero, and therefore
\begin{equation}
\nabla_\mu u^\mu = \frac{\gamma}{t} \left[ 2\frac{v}{\xi} + \left[1 - \gamma^2 v (\xi - v) \right] \frac{\partial v}{\partial \xi} \right]
\end{equation}
The gradients in eq.~\eqref{eq:continuity1} and~\eqref{eq:continuity2} can also be expressed as~\cite[eq. 25]{espinosa_energy_2010}.
\begin{align}
u_\mu \nabla^\mu &= - \frac{\gamma}{t} (\xi - v) \frac{\partial}{\partial \xi}, \\
\bar{u}_\mu \nabla^\mu &= \frac{\gamma}{t} (1 - \xi v) \frac{\partial}{\partial \xi}.
\end{align}
With these eq.~\eqref{eq:continuity1} and~\eqref{eq:continuity2} can be expressed as
\begin{align}
(\xi - v) \frac{1}{w} \frac{de}{d\xi} &= 2 \frac{v}{\xi} + \left[ 1 - \gamma^2 v (\xi - v) \right] \frac{dv}{d\xi}, \\
(1 - v\xi) \frac{d_\xi p}{w} &= \gamma^2 (\xi - v) \frac{dv}{d\xi}.
\end{align}
These are dependent only on $\xi$ instead of $r$,
and therefore a spherically symmetric self-similar solution exists.
Here the derivatives are now total derivatives, as $e$ and $p$ are functions of a single variable.
By dividing one equation by the other and using eq.~\eqref{eq:cs2_compact}, we can combine these to
\begin{equation}
2 \frac{v}{\xi} = \gamma^2 (1 - v\xi) \left[ \frac{\mu^2}{c_s^2} - 1 \right] \frac{dv}{d\xi},
\label{eq:continuity_combined}
\end{equation}
where
\begin{equation}
\mu(\xi,v) \equiv \frac{\xi - v}{1 - \xi v}
\label{eq:mu}
\end{equation}
is the
Lorentz transformed fluid velocity
$\xi$ in a frame moving outward with the speed $v$.
This can be rearranged to give
\cites[eq. 7.30-7.31]{lecture_notes}[eq. 5]{giese_2021}
\begin{equation}
\frac{dv}{d\xi} = \frac{2v(1-v^2)}{\xi(1-v\xi)} \left( \frac{\mu(\xi,v)^2}{c_s^2} - 1 \right)^{-1}.
\label{eq:hydro_diff1}
\end{equation}
From eq.~\eqref{eq:continuity2} using eq.~\eqref{eq:cs2_compact} and~\eqref{eq:enthalpy_sum} we also get
\begin{equation}
\frac{dw}{d\xi} = w \left( 1 + \frac{1}{c_s^2} \right) \gamma^2 \mu(\xi,v) \frac{dv}{d\xi}.
\label{eq:hydro_diff2}
\end{equation}

Equation~\eqref{eq:hydro_diff1} can be split in two using an auxiliary quantity $\tau$,
resulting in the equation group
\cite[eq. B.14-16]{hindmarsh_gw_pt_2019}
\begin{align}
\frac{d\xi}{d\tau} &= \xi \left[ (\xi - v)^2 - c_s^2 (1 - \xi v)^2 \right],
\label{eq:hydro_param1} \\
\frac{dv}{d\tau} &= 2 v c_s^2 (1 - v^2) (1 - \xi v),
\label{eq:hydro_param2} \\
\frac{dw}{d\tau} &= w \left( 1 + \frac{1}{c_s^2} \right) \gamma^2 \mu \frac{dv}{d\tau}.
\label{eq:hydro_param3}
\end{align}
This is easier to prove in the reverse direction by dividing one equation by the other.
It should be noted that in general $c_s^2 = c_s^2(w,\phi)$, which tightens the coupling between these equations.
The parametric equations have fixed points at
$(\xi,v) = (0,0)$ and
$(\xi,v) = (1,1)$.
When $c_s$ is a constant, the equations also have a fixed point at
$(\xi,v) = (c_s,0)$.
When $c_s$ is constant, the first two equations are not dependent on enthalpy, which simplifies the integration considerably.
However, in the general case of $c_s(w,\phi)$ the equations have to be integrated together.

\section{Types of solutions}
\label{solution_types}
Solutions to the hydrodynamic equations~\eqref{eq:hydro_param1}, and~\eqref{eq:hydro_param2} are obtained by integrating away from the position of the bubble wall, $\xi_w$.
There are two relevant external boundary conditions.
To maintain spherical symmetry
\begin{equation}
\lim_{\xi \rightarrow 0} v = 0.
\end{equation}
To maintain causality
\begin{equation}
\lim_{\xi \rightarrow 1} v = 0,
\end{equation}
as no information can propagate faster than light, and we assume the fluid to be stationary until a signal from the expanding bubble arrives.
In addition to these there is one boundary condition for each side of the wall as
\begin{equation}
\lim_{\xi \rightarrow \xi_w^\pm = v_w \pm \delta, \ \delta \rightarrow 0} v = v_\pm = \mu (\xi_w, \tilde{v}_\pm),
\end{equation}

There are two ways to satisfy these conditions.
We can either start at $v=0$, or from a region where $\xi > c_{s-}(w)$ and $\mu(\xi,v) > c_{s-}(w)$, and therefore $\frac{dv}{d\xi} > 0$,
and integrate backwards in $\xi$ over a route where $\mu(\xi,v) > c_{s-}(w)$ is satisfied.
The other way to reach $v=0$ is by a discontinuity, i.e. a shock.
This leads to two classes of solutions.
Another way to see that there are two classes of solutions is by noting
that the equations~\eqref{eq:v_tilde_plus} and~\eqref{eq:v_tilde_minus} have two branches.
These solutions are classified in table~\ref{table:solution_types}.

\begin{table}[ht!]
\small
\caption{Types of solutions to the hydrodynamic equations}
\begin{tabular}{r|c|c}
                & Detonations            & Deflagrations \\
                & $p_+ < p_-, \tilde{v}_+ > \tilde{v}_-$ & $p_+ > p_-, \tilde{v}_+ < \tilde{v}_-$ \\ \hline
Weak            & $\tilde{v}_+ > c_{s+}(w_+), \ \tilde{v}_- > c_{s-}(w_-)$ & $\tilde{v}_+ < c_{s+}(w_+), \ \tilde{v}_- < c_{s-}(w_-)$ \\
Chapman-Jouguet & $\tilde{v}_+ > c_{s+}(w_+), \ \tilde{v}_- = c_{s-}(w_-)$ & $\tilde{v}_+ < c_{s+}(w_+), \ \tilde{v}_- = c_{s-}(w_-)$ \\
Strong          & {\color{gray} $\tilde{v}_+ > c_{s+}(w_+), \ \tilde{v}_- < c_{s-}(w_-)$} & {\color{gray} $\tilde{v}_+ < c_{s+}(w_+), \ \tilde{v}_- > c_{s-}(w_-)$} \\
\end{tabular}
\label{table:solution_types}
\end{table}

The solutions with $\tilde{v}_+ > \tilde{v}_-$ are known as detonations.
Detonations are further characterised by the scale of their $\tilde{v}_-$.
In weak detonations $\tilde{v}_- > c_{s-}(w_-)$.%
\footnote{The note in \cite[p. 265]{rezzolla_relativistic_2013} that weak detonations are not possible in an exothermic reaction does not apply to cosmological phase transitions, since in them the number of particles is not conserved.}
Correspondingly, detonations with $\tilde{v}_- < c_{s-}(w_-)$ are known as strong detonations.
However, they are unstable and will naturally evolve into the third class of detonations,
the Chapman-Jouguet detonations, for which $\tilde{v}_- = c_{s-}(w_-)$.
\cite[p. 279]{rezzolla_relativistic_2013}
Since we are investigating a self-similar bubble that has already evolved for some time,
the most of our detonations are weak detonations,
and those at the detonation-hybrid boundary are Chapman-Jouguet detonations.
In detonations the shock and phase boundary fronts are unified to a single front.
\cite{kurki-suonio_supersonic_1995}

The solutions where $\tilde{v}_+ < \tilde{v}_-$ are known as deflagrations.
In them the phase front can influence the fluid ahead, and the wall is preceded by an accelerating fluid and a shock.
Strong deflagrations are not possible in an exothermic reaction.
\cite[p. 267]{rezzolla_relativistic_2013}
Therefore, only weak and Chapman-Jouguet deflagrations are possible.
In weak detonations the fluid inside the phase boundary is still, and the preceding shock is weak and
known as the precompression front.
\cite{rezzolla_relativistic_2013}

\begin{figure}[ht!]
\centering
\includegraphics[width=\textwidth]{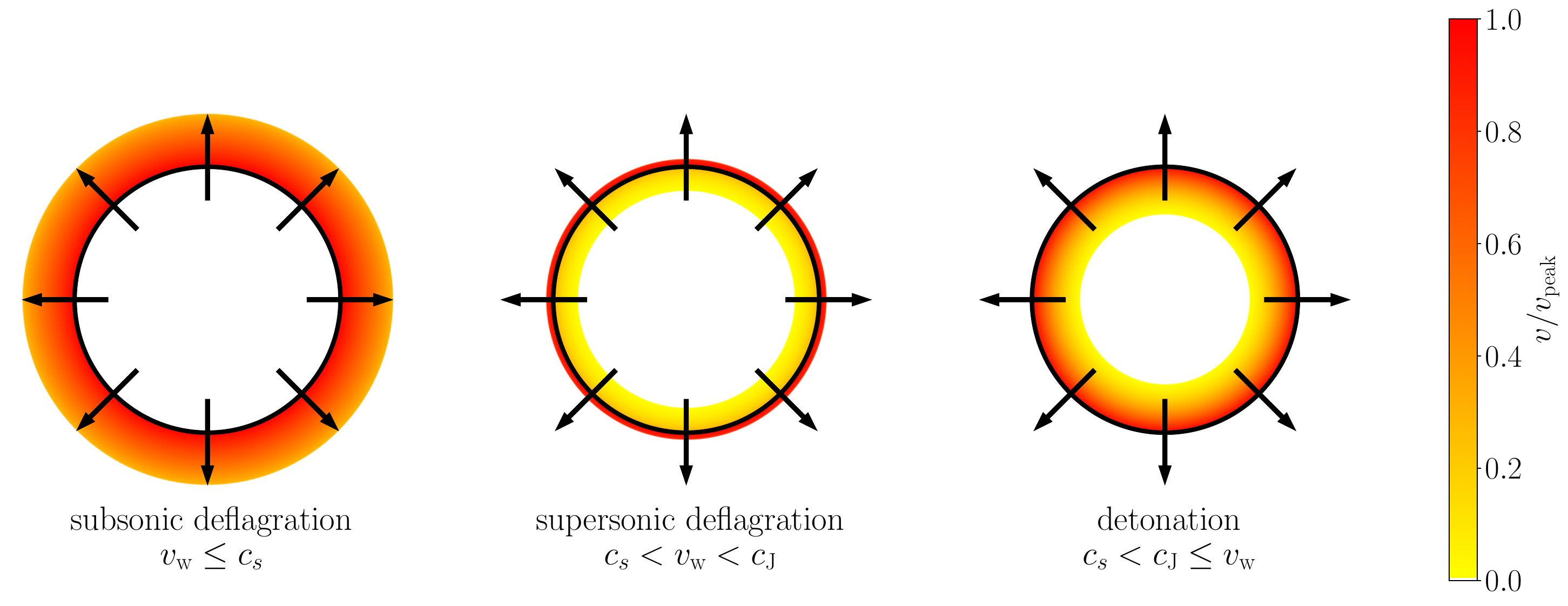}
\caption{Three different types of relativistic combustion. Generated with PTtools. See also \cites[fig. 14]{lecture_notes}[fig. 14]{mazumdar_review_2019}.}
\label{fig:solution_types}
\end{figure}

The third class of solutions are the supersonic deflagrations, also known as hybrids.
They consist of a Chapman-Jouguet deflagration followed by a detonation-like rarefaction wave.
In hybrids the wall speed exceeds the speed of sound in the broken phase $c_{s-}$,
and the fluid is moving inside the phase boundary as well, as in a detonation.
\cites{kurki-suonio_supersonic_1995}[p. 37]{lecture_notes}[p. 35]{hindmarsh_gw_pt_2019}
The three solution types: subsonic deflagrations, supersonic deflagrations, aka. hybrids, and detonations are illustrated in figure~\ref{fig:solution_types}.

\clearpage
\FloatBarrier
\subsection{Speed limits}
The observables of a physical system must be real, including $\tilde{v}_+$ and $\tilde{v}_-$.
Therefore, the expressions in the square roots of eq.~\eqref{eq:v_tilde_plus} and~\eqref{eq:v_tilde_minus} must be real.
By simplifying the expression in the square root of eq.~\eqref{eq:v_tilde_plus} we see that the square root is real $\forall \alpha_+ \geq 0$,
but setting the square root in eq.~\eqref{eq:v_tilde_minus} to zero gives a limit for $\tilde{v}_+$ as
\begin{equation}
\tilde{v}_+ = \frac{1}{\sqrt{3}} \left( \frac{1 \pm \sqrt{2 \alpha_+ + 3 \alpha_+^2}}{1 + \alpha_+} \right).
\label{eq:v_tilde_plus_limit}
\end{equation}
The positive sign is the lower limit for detonations, and the negative sign is the upper limit for deflagrations.%
\footnote{This is the same equation as in \cites[eq. 7.34]{lecture_notes}[eq. B.19]{hindmarsh_gw_pt_2019},
but in those articles there is a typo due to which a factor of 2 is missing from the expression.}

Inserting this value to the equation~\eqref{eq:v_tilde_minus} and simplifying the expression with the knowledge that the square root is now zero
we get that
\begin{equation}
\tilde{v}_- = \frac{1}{\sqrt{3}}.
\label{eq:v_tilde_minus_limit}
\end{equation}
This is the lower limit for detonations, and the upper limit for deflagrations.

The Chapman-Jouguet speed is defined as
\begin{equation}
\tilde{v}_+=v_{CJ} \Leftrightarrow \tilde{v}_- = c_{s-}(w_-).
\label{eq:chapman_jouguet}
\end{equation}
Starting from section~\eqref{bag_model} we will see that for some models the speed limit set by the Chapman-Jouguet speed of~\eqref{eq:chapman_jouguet}
and the condition of~\eqref{eq:v_tilde_minus_limit} that the observables are real are equivalent,
but in general this is not the case.

Now we have the necessary knowledge to classify the different regions of fig.~\eqref{fig:vplus_vminus}.
If the speeds of sound are equivalent to $\frac{1}{\sqrt{3}}$,
we can have only weak and Chapman-Jouguet detonations and deflagrations.
When the speed of sound in the broken phase $c_{s-}(w_-) < \frac{1}{\sqrt{3}}$,
we could have a strong deflagration where $c_{s-}(w_-) < \tilde{v}_- < \frac{1}{\sqrt{3}}$,
but this is ruled out by the fact that the phase transition is exothermic.
Correspondingly, when $c_{s-}(w_-) > \frac{1}{\sqrt{3}}$,
we can have a strong detonation where $\frac{1}{\sqrt{3}} < \tilde{v}_-(w_-, \phi_-) < c_{s-}(w_-)$,
but as previously discussed, a strong detonation is unstable and will evolve into a Chapman-Jouguet detonation.
Since for Chapman-Jouguet detonations $v_w = \tilde{v}_+ > \tilde{v}_-$,
the Chapman-Jouguet speed is the lower speed limit for detonations that have been evolving for a sufficiently long time.

\section{General equation of state}
\label{general_eq}
The equation of state is the function $p(T)$ or $p(w)$ that characterises the behaviour of the fluid as a function of temperature $T$ or enthalpy $w$.
It should be highlighted that everything up to this point has been independent of the choice of the equation of state.
We will now first describe the derivation for the general equation of state of an ultrarelativistic plasma,
and then introduce approximate models.

The energy density at a point is the sum of the energies of all particles at that point.
Therefore, the energy density in the position space can be obtained by integrating over the momentum space as
\cite[eq. 4.10]{lecture_notes}
\begin{equation}
e(x) = \int \frac{d^3 p}{(2 \pi)^3} E(p) f(\vec{p}, x).
\end{equation}
Similarly we can obtain the other components of the energy-momentum tensor in the position space as
\cite[eq. 4.13]{lecture_notes}
\begin{equation}
T^{\mu \nu}(x) = \int \frac{d^3 p}{(2 \pi)^3} \frac{p^\mu p^\nu}{p^0} f(\vec{p},x).
\end{equation}
To give an intuitive explanation for this we can note that $p^0 = E = \gamma m_0$ and $p^i = \gamma m_0 v^i$.
Therefore, $\frac{p^i}{p^0} = v^i$,
and the factor $\frac{p^i p^j}{p^0} = p^i v^j$ denotes the amount of momentum being transported by the moving plasma.
By inserting this to the definition of pressure~\eqref{eq:pressure_from_ep_tensor} we get
\begin{equation}
p(x) = \frac{1}{3} \int \frac{d^3 p}{(2 \pi)^3} \frac{|\vec{p}|^2}{E} f(\vec{p},x)
\end{equation}
This notation is rather confusing, as the $p$ on the left refers to the pressure,
and all $p$ on the right refer to the momentum.

The particle distribution functions for fermions and bosons are
\cite[eq. 4.6]{lecture_notes}
\begin{equation}
f(\vec{p}) = \frac{1}{e^\frac{E-\mu}{T} \pm 1},
\end{equation}
where $+$ gives the Fermi-Dirac distribution for fermions and $-$ gives the Bose-Einstein distribution for bosons.
Let us approximate that the fluid consists entirely of bosons,
and that they are ultrarelativistic: $E(\vec{p}) = \sqrt{\vec{p}^2 + m^2} \approx p$.
Particle number is not conserved in an ultrarelativistic plasma, and therefore $\mu = 0$.
Let us also remember that we can convert the integral to 1D using the area of a sphere $A(r) = 4\pi r^2$.
With these we get
\begin{align}
e &= \frac{1}{2 \pi^2} \int_0^\infty \frac{p^3 dp}{e^\frac{p}{T} - 1},
\label{eq:ultrarelativistic_e} \\
p &= \frac{1}{6 \pi^2} \int_0^\infty \frac{p^3 dp}{e^\frac{p}{T} - 1}.
\label{eq:ultrarelativistic_p}
\end{align}
These contain integrals of the form
\cite[eq. B.36]{schroeder_thermal_2000}
\begin{equation}
\int_0^\infty \frac{x^n}{e^x - 1} = \Gamma(n+1) \zeta(n+1),
\end{equation}
where $\Gamma$ is the
gamma function, which for integers is $\Gamma(n) = (n-1)!$, and $\zeta$ is the Riemann zeta function.
It can be shown that $\zeta(4) = \frac{\pi^4}{90}$.
\cite[prob. B.19]{schroeder_thermal_2000}
At this ultrarelativistic limit we can see from eq.~\eqref{eq:ultrarelativistic_e} and~\eqref{eq:ultrarelativistic_p} that $e = 3p$, and therefore using eq.~\eqref{eq:cs2_compact} that $c_s^2 = \frac{1}{3}$.

To account for the various particle species in the plasma and their non-ultrarelativistic behaviour,
we need to multiply with the degrees of freedom $g_e$ and $g_p$,
which are weighted sums over the contributions of different particle species.
To also account for the field, we need to add its potential $V$ to the final result
\cite[eq. S12]{borsanyi_lattice_2016}
\begin{align}
e(T,\phi) &= \frac{\pi^2}{30} g_e(T) T^4 + V(T,\phi),
\label{eq:e_general} \\
p(T,\phi) &= \frac{\pi^2}{90} g_p(T) T^4 - V(T,\phi).
\label{eq:p_general}
\end{align}
This means that we assume that all particles get their masses from the potential.

With eq.~\eqref{eq:enthalpy_sum},~\eqref{eq:enthalpy_entropy},~\eqref{eq:e_general} and~\eqref{eq:p_general} we can obtain the entropy density as
\cite[eq. S12]{borsanyi_lattice_2016}
\begin{equation}
s(T,\phi) = \frac{2\pi^2}{45} g_s(T) T^3,
\label{eq:s_general}
\end{equation}
where the degrees of freedom for the entropy density are
\begin{equation}
g_s = \frac{1}{4} (3g_e + g_p).
\end{equation}
Often only $g_e$ and $g_s$ are provided, and $g_p$ is obtained with
\begin{equation}
g_p = 4g_s - 3g_e.
\end{equation}
For a free scalar particle $g_e = g_p = g_s = 1$, which is known as the Stefan-Boltzmann limit.

\section{The bag model}
\label{bag_model}
For phase transitions in the early universe,
the most commonly used equation of state is the bag model, for which
\begin{equation}
g_{e\pm} = g_{p\pm} = g_{s\pm} = \frac{90}{\pi^2} a_\pm,
\end{equation}
where $a_\pm$ are constants with $a_+ > a_-$.
$+$ refers to the symmetric phase, and $-$ refers to the broken phase.
The potentials $V_\pm$ are also constants with $V_+ > V_-$.
This results in the bag equation of state
\cites[eq. 7.33]{lecture_notes}[eq. 8-9]{giese_2020}
\begin{equation}
p_\pm = a_\pm T^4 - V_\pm.
\label{eq:bag_p}
\end{equation}
The potential difference $\Delta V \equiv V_+ - V_-$, also known as $\epsilon$,
is the temperature-independent vacuum energy that is released in the phase transition.
The potentials are usually shifted so that $V_b = 0$.
The name of the bag model originates from quantum chromodynamics (QCD),
where it's used to describe the proton as a bag of quarks.
\cite{giese_2020}

Using eq.~\eqref{eq:entropy_density} we get the entropy density
\begin{equation}
s_\pm = 4 a_\pm T^3,
\end{equation}
and with eq.~\eqref{eq:enthalpy_entropy} the enthalpy density
\begin{equation}
w_\pm = 4 a_\pm T^4.
\end{equation}
Finally with eq.~\eqref{eq:enthalpy_sum} we get the energy density
\begin{equation}
e_\pm = 3 a_\pm T^4 + V_\pm.
\end{equation}
The speed of sound from eq.~\eqref{eq:cs2_explicit} simplifies to
\begin{equation}
c_s^2 = \frac{dp}{de} = \frac{dp/dT}{de/dT} = \frac{1}{3}.
\end{equation}
As in section~\ref{general_eq}, this corresponds to assuming both of the phases to be ultrarelativistic.
It also happens to be the same as the $\tilde{v}_-$ that corresponds to eq.~\eqref{eq:v_tilde_plus_limit},
and therefore the Chapman-Jouguet speed $v_{CJ}$ of the bag model is given by eq.~\eqref{eq:v_tilde_plus_limit}.
The trace anomaly of eq.~\eqref{eq:theta} simplifies to
\begin{align}
\theta_\pm = V_\pm.
\end{align}
Therefore, the transition strength of~\eqref{eq:alpha_plus} simplifies to
\begin{equation}
\alpha_{+,\text{bag}} = \frac{4 \Delta V}{3 w_+},
\label{eq:alpha_plus_bag}
\end{equation}
and the transition strength at nucleation temperature from~\eqref{eq:alpha_n} simplifies to
\begin{equation}
\alpha_{n,\text{bag}} = \frac{4 \Delta V}{3 w_n}.
\label{eq:alpha_n_bag}
\end{equation}
This is trivial to invert, giving enthalpy at the nucleation temperature $w_n$.

\section{The constant sound speed model}
\label{const_cs}
The constant sound speed model, also known as the $\mu_\pm$ model, the $\mu, \nu$ model,
or in some sources as the template model,
expands the bag model by allowing a different speed of sound in each phase.
\cites{leitao_hydrodynamics_2015}{giese_2020}{giese_2021}
It is obtained by setting
\begin{equation}
g_{p\pm} = \frac{90}{\pi^2} a_\pm \left( \frac{T}{T_0} \right)^{\mu_\pm - 4},
\end{equation}
where $\mu_\pm$ are constants.
This results in the equation of state
\cites[eq. 15]{giese_2021}[eq. 38]{giese_2020}%
\footnote{Please note that there is a typo in \cite[eq. 15]{giese_2021}. There the 4 should be a $\mu$.}
\begin{equation}
p_\pm = a_\pm \left( \frac{T}{T_0} \right)^{\mu_\pm - 4} T^4 - V_\pm.
\end{equation}
The reference temperature $T_0$ is introduced for the consistency of the units.
For simplicity it can be set to 1 GeV, or whichever unit one is using for the temperature.
Several articles use a version with $T^{\mu_\pm}$, which is easier to read but has inconsistent units.
The parameters $\mu_\pm$ are determined by the sound speeds in the symmetric and broken phases with
\cites[eq. 16]{giese_2021}[eq. 39]{giese_2020}
\begin{equation}
\mu_\pm \equiv 1 + \frac{1}{c_{s\pm}^2}.
\end{equation}
This model reduces to the bag model,
when $c_{s+}^2 = c_{s-}^2 = \frac{1}{3}$
and equivalently
$\mu_+ = \mu_- = 4$.

Similarly as for the bag model we can derive the entropy density
\begin{equation}
s_\pm = \mu_\pm a_\pm \left( \frac{T}{T_0} \right)^{\mu_\pm - 4} T^3,
\end{equation}
and from it the enthalpy density
\begin{equation}
w_\pm = \mu_\pm a_\pm \left( \frac{T}{T_0} \right)^{\mu_\pm - 4} T^4,
\label{eq:w_const_cs}
\end{equation}
and finally the energy density
\begin{equation}
e_\pm = (\mu_\pm - 1) a_\pm \left( \frac{T}{T_0} \right)^{\mu_\pm - 4} T^4 + V_\pm.
\end{equation}
The equation for enthalpy density~\eqref{eq:w_const_cs} can be inverted to give the temperature as a function of enthalpy
\begin{equation}
T_\pm = \left( \frac{w}{\mu_\pm a_\pm T_0^4} \right)^\frac{1}{\mu_\pm} T_0.
\end{equation}
Therefore the effective degrees of freedom are
\begin{align}
g_{p\pm} &= \frac{90}{\pi^2} a_\pm \left( \frac{T}{T_0} \right)^{\mu_\pm - 4}, \\
g_{s\pm} &= \frac{45}{2\pi^2} \mu_\pm a_\pm \left( \frac{T}{T_0} \right)^{\mu_\pm - 4}, \\
g_{e\pm} &= \frac{30}{\pi^2} (\mu - 1) a_\pm \left( \frac{T}{T_0} \right)^{\mu_\pm - 4}.
\end{align}

The trace anomaly of eq.~\eqref{eq:theta} simplifies to
\begin{align}
\theta_\pm(w)
&= \left( \frac{\mu_\pm}{4} - 1 \right) a_\pm \left( \frac{T}{T_0} \right)^{\mu_\pm - 4} T^4 + V_\pm \\
&= \left( \frac{1}{4} - \frac{1}{\mu_\pm} \right) w_\pm + V_\pm.
\label{eq:theta_const_cs}
\end{align}
The transition strength of eq.~\eqref{eq:alpha_plus} simplifies to
\begin{equation}
\alpha_+ = \frac{1}{3} \left( 1 - \frac{4}{\mu_+} \right) - \frac{1}{3} \left(1 - \frac{4}{\mu_-} \right) \frac{w_-}{w_+} + \alpha_{+,\text{bag}}.
\label{eq:alpha_plus_const_cs}
\end{equation}
The transition strength at nucleation temperature of eq.~\eqref{eq:alpha_n} is correspondingly
\begin{align}
\alpha_n &= \frac{1}{3} \left( 1 - \frac{4}{\mu_+} \right) - \frac{1}{3} \left(1 - \frac{4}{\mu_-} \right) \frac{w(T_n, \phi_-)}{w_n} + \alpha_{n,\text{bag}} \\
&= \frac{4}{3} \left( \frac{1}{\mu_-} - \frac{1}{\mu_+} \right) + \frac{1}{3} \left( 1 - \frac{4}{\mu_-} \right)
\left( 1 - \frac{\mu_- a_-}{\mu_+ a_+} \left( \frac{T_n}{T_0} \right)^{\mu_- - \mu_+} \right) + \alpha_{n,\text{bag}}.
\label{eq:alpha_n_const_cs}
\end{align}
When $\mu_- = 4$, $\alpha_n$ of eq.~\eqref{eq:alpha_n_const_cs} becomes independent of $w_-$, resulting in
\begin{equation}
w_n = \frac{4 (V_+ - V_-)}{3 \alpha_n + \frac{4}{\mu_+} - 1}.
\label{eq:wn_const_cs_mu4}
\end{equation}
In this case we can obtain $v_{CJ}(\alpha_n, c_{s+})$ analytically.
First we insert $\alpha_n$ to eq.~\eqref{eq:wn_const_cs_mu4} to get $w_n$,
and then insert $w_n$ to eq.~\eqref{eq:alpha_plus_const_cs} to get $\alpha_+$,
since in this case $\alpha_+$ is independent of $w_-$.
Then we can insert $c_{s-}$ and $\alpha_+$ to eq.~\eqref{eq:v_tilde_plus} to get $\tilde{v}_+ = v_{CJ}$.

The pseudotrace of eq.~\eqref{eq:theta_bar} is given by
\begin{align}
\bar{\Theta} = (\mu_\pm - \mu_-) a \left(\frac{T}{T_0}\right)^{\mu_\pm - 4} T_0^4 + \mu_- V_\pm.
\end{align}
Therefore the corresponding transition strengths of eq.~\eqref{eq:alpha_theta_bar_plus} and~\eqref{eq:alpha_theta_bar_n} become
\begin{align}
\alpha_{\bar{\Theta}_+} &= \frac{1}{3} \left(1 - \frac{\mu_-}{\mu_+}\right) + \frac{\mu_-}{4} \alpha_{+,\text{bag}}, \\
\alpha_{\bar{\Theta}_n} &= \frac{1}{3} \left(1 - \frac{\mu_-}{\mu_+}\right) + \frac{\mu_-}{4} \alpha_{n,\text{bag}}.
\end{align}
These are equal for detonations regardless of the wall speed or the speeds of sound,
since $\alpha_{+,\text{bag}} = \alpha_{n,\text{bag}}$ for detonations.

If we set $T_0 = T_c$, eq.~\eqref{eq:critical_temp} results in
\begin{equation}
\Delta V = (a_+ - a_-) T_c^4.
\end{equation}
Using this the ``bag'' part in eq.~\eqref{eq:alpha_plus_const_cs} and~\eqref{eq:alpha_n_const_cs} becomes
\begin{align}
\alpha_{+,\text{bag}} &= \frac{4}{3 \mu_+} \left( 1 - \frac{a_-}{a_+} \right) \left( \frac{T_+}{T_c} \right)^{-\mu_+}, \\
\alpha_{n,\text{bag}} &= \frac{4}{3 \mu_+} \left( 1 - \frac{a_-}{a_+} \right) \left( \frac{T_n}{T_c} \right)^{-\mu_+}.
\end{align}
Using this we can reorder eq.~\eqref{eq:alpha_n_const_cs} to
\begin{equation}
\frac{a_-}{a_+} = \frac{4 + \left( \mu_+ - 4 - 3 \mu_+ \alpha_n \right) \left(\frac{T_n}{T_c}\right)^{\mu_+}}{4 + \left( \mu_- - 4 \right) \left(\frac{T_n}{T_c}\right)^{\mu_-}}.
\end{equation}

Approximating $T_+ \approx T_-$ results in eq.~\eqref{eq:cs2_compact} being approximated as
\begin{equation}
c_{s,b}^2
\equiv \frac{dp_b}{de_b}
= \frac{dp_b/dT}{de_b/dT} \big|_{T_+ \approx T_-}
\approx \frac{\delta p_-}{\delta e_-}.
\label{eq:const_cs_approx}
\end{equation}
In the case of the constant sound speed model, $c_{s}$ is independent of temperature, and therefore this approximation is exact.
With this approximation and the use of eq.~\eqref{eq:delta_relation} the second junction condition of eq.~\eqref{eq:junction_conditions_simplified} becomes
\begin{equation}
\delta p_- \left(1 - \frac{\tilde{v}_+ \tilde{v}_-}{c_{s,b}^2} \right) = \tilde{v}_+ \tilde{v}_- De_+ - Dp_+.
\end{equation}
Using this, eq.~\eqref{eq:delta_relation} and eq.~\eqref{eq:enthalpy_sum} on the first junction condition of~\eqref{eq:junction_conditions_simplified} results in
\begin{align}
\frac{\tilde{v}_+}{\tilde{v}_-}
&\approx \frac{
w_+ \left( \frac{\tilde{v}_+ \tilde{v}_-}{c_{s,b}^2} - 1 \right) + \left( De_+ - \frac{Dp_+}{c_{s,b}^2} \right)
}{
w_+ \left( \frac{\tilde{v}_+ \tilde{v}_-}{c_{s,b}^2} - 1 \right) + \tilde{v}_+ \tilde{v}_- \left( De_+ - \frac{Dp_+}{c_{s,b}^2} \right)
} \\
&= \frac{ \left( \frac{\tilde{v}_+ \tilde{v}_-}{c_{s,b}^2} - 1 \right) + 3 \alpha_{\bar{\Theta}_+}
}{
\left( \frac{\tilde{v}_+ \tilde{v}_-}{c_{s,b}^2} - 1 \right) + 3\tilde{v}_+ \tilde{v}_- \alpha_{\bar{\Theta}_+}}.
\label{eq:const_cs_vp_vm}
\end{align}
The only approximation in this calculation is that of eq.~\eqref{eq:const_cs_approx}.
As it is exact for the constant sound speed model,
eq.~\eqref{eq:const_cs_vp_vm} is exact for the constant sound speed model.
Eq.~\eqref{eq:const_cs_vp_vm} can be rearranged to
\begin{equation}
\frac{\tilde{v}_+}{c_{s,b}^2} \tilde{v}_-^2
+ \left(3 \alpha_{\bar{\Theta}_+} - 1 - \tilde{v}_+^2 \left(\frac{1}{c_{s,b}^2} + 3 \alpha_{\bar{\Theta}_+} \right) \right) \tilde{v}_-
+ \tilde{v}_+
= 0,
\end{equation}
which is a basic second-order equation for $\tilde{v}_- ( \tilde{v}_+, \alpha_{\bar{\Theta}_+} )$.
Since $\tilde{v}_+ = v_w$ for all detonations, and $\alpha_{\bar{\Theta}_+} = \alpha_{\bar{\Theta}_n}$ for all detonations in the constant sound speed model, this gives $\tilde{v}_- ( v_w, \alpha_{\bar{\Theta}_n} )$.
Eq.~\eqref{eq:const_cs_vp_vm} can also be rearranged to
\begin{equation}
\left( \frac{1}{c_{s,b}^2} + 3\alpha_{\bar{\Theta}_+} \right) \tilde{v}_+^2
- \left( \frac{1}{\tilde{v}_-} + \frac{\tilde{v}_-}{c_{s,b}^2} \right) \tilde{v}_+
+ 1 - 3\alpha_{\bar{\Theta}_+}
= 0.
\end{equation}
This gives the Chapman-Jouguet speed $\tilde{v}_+ \left( \tilde{v}_- = c_{s,b} \right)$
as~\cite[eq. 55]{giese_2020}
\begin{equation}
v_{CJ} = \frac{ 1 + \sqrt{ 3\alpha_{\bar{\Theta}_+} ( 1 - c_{s,b}^2 + 3 c_{s,b}^2 \alpha_{\bar{\Theta}_+} ) }}{ \frac{1}{c_{s,b}} + 3 c_{s,b} \alpha_{\bar{\Theta}_+}}.
\end{equation}
This holds for all values of $\mu_+$ and $\mu_-$.

\clearpage
\section{Energy redistribution}
\label{energy_redistribution}
Once the velocity profile of a bubble is known, it can be used to compute various thermodynamic quantities.

We have assumed that the background spacetime is Minkowski,
and that the spacetime does not expand significantly during the phase transition.
Therefore, the total energy of the system is conserved, and
\begin{equation}
E = 4 \pi \int_0^R dr r^2 T^{00}
\end{equation}
is a constant for $R$ larger than the fluid velocity profile.
It should be noted that this energy conservation applies only to the total energy of the system,
and not the energy of the fluid, as there is energy transfer from the field to the fluid.
\cite[p. 21]{lecture_notes}
Correspondingly the mean energy density $\bar{e}$ is defined as~\cite[p. 39]{lecture_notes}
\begin{equation}
\bar{e} = \frac{3}{\xi_\text{max}^3} \int_0^{\xi_\text{max}} d\xi \xi^2 e = e(w_n, \phi_s).
\label{eq:e_conservation}
\end{equation}
The mean enthalpy density $\bar{w}$ is defined as
\begin{equation}
\bar{w} = \frac{3}{\xi_\text{max}^3} \int_0^{\xi_\text{max}} d\xi \xi^2 w.
\label{eq:wbar}
\end{equation}
Using these we can define the energy fluctuation variable
\begin{equation}
\lambda(x) \equiv \frac{e(x) - \bar{e}}{\bar{w}}.
\label{eq:lambda}
\end{equation}

For an ideal fluid of eq.~\eqref{eq:ep_tensor}~\cite[eq. B.23]{hindmarsh_gw_pt_2019}
\begin{equation}
T^{00} = w\gamma^2 - p = w\gamma^2 v^2 + e = w\gamma^2 v^2 + \frac{3}{4}w + \theta.
\end{equation}
Inserting this to the energy conservation equation~\eqref{eq:e_conservation}
results in the equation~\cite[eq. B.24]{hindmarsh_gw_pt_2019}
\begin{equation}
e_K + \Delta e_Q = - \Delta e_\theta,
\label{eq:energy_components}
\end{equation}
which consists of the terms defined below.
The kinetic energy density is given by
\begin{equation}
e_K \equiv 4 \pi \int_0^{\xi_\text{max}} d\xi \xi^2 w \gamma^2 v^2.
\label{eq:kinetic_energy_density}
\end{equation}
It denotes the energy that is converted to macroscopic fluid movement.
The thermal energy density difference is given by
\begin{equation}
\Delta e_Q \equiv 4 \pi \int_0^{\xi_\text{max}} d\xi \xi^2 \frac{3}{4} (w - w_n).
\label{eq:thermal_energy_density}
\end{equation}
It denotes the energy that is converted to microscopic fluid movement.
The trace anomaly difference is given by
\begin{equation}
\Delta e_\theta \equiv 4 \pi \int_0^{\xi_\text{max}} d\xi \xi^2 (\theta - \theta_n).
\end{equation}
It can be considered as the potential energy available for transformation.
We can obtain the volume-averaged entropy density difference similarly as the energy density, with
\begin{equation}
\Delta s_\text{avg} = 4\pi \int_0^{\xi_\text{max}} d\xi \xi^2 \left( s(w,\phi) - s(w_n, \phi_s) \right).
\end{equation}
The upper integration limit $\xi_\text{max}$ can be chosen arbitrarily as long as it's outside the bubble,
as outside the bubble all three quantities go to zero outside the fluid shell due to $v=0$, $w=w_n$ and $\theta = \theta_n$.
A convenient choice is $\xi_\text{max} = \max (v_w, \xi_{sh})$.
\cite[eq. B.25]{hindmarsh_gw_pt_2019}

It should be noted that the trace anomaly is not quite equivalent to the thermal potential energy density $V_T(T,\phi)$,
as using eq.~\eqref{eq:enthalpy_pressure},~\eqref{eq:enthalpy_sum} and~\eqref{eq:p_general}
we can see that for the case of temperature-independent $g$,
\begin{equation}
\theta = V_T - \frac{1}{4} T \frac{\partial V_T}{\partial T}.
\end{equation}
Therefore, not all of the potential energy difference can be turned into kinetic and thermal energy.
\cite[ch. B.2]{hindmarsh_gw_pt_2019}

The fraction of total energy that is converted to kinetic energy is known as the kinetic energy fraction
\begin{equation}
K \equiv \frac{e_K}{\bar{e}}.
\label{eq:kinetic_energy_fraction}
\end{equation}

Kinetic and thermal efficiency factors quantify the fraction of the available energy $\Delta e_\theta$
that is converted to kinetic and thermal energy.
They can be defined as
\begin{equation}
\kappa \equiv \frac{e_K}{| \Delta e_\theta |}, \quad
\omega \equiv \frac{\Delta e_Q}{\Delta e_\theta}.
\label{eq:kappa_omega}
\end{equation}
Since energy is conserved by eq.~\eqref{eq:energy_components},
these have have the property that $\kappa + \omega = 1$.

It should be noted that some sources such as~\cites[eq. 36]{giese_2020}[eq. 12, 14]{giese_2021}
use different definitions for $\kappa$:
\begin{align}
\kappa_{\bar{\Theta}_+} &\equiv \frac{4 e_K}{D \bar{\Theta}(T_+)} = \frac{4 e_K}{3 \alpha_{\bar{\Theta}_+} w_+},
\label{eq:kappa_thetabar_plus} \\
\kappa_{\bar{\Theta}_n} &\equiv \frac{4 e_K}{D \bar{\Theta}(T_n)} = \frac{4 e_K}{3 \alpha_{\bar{\Theta}_n} w_n}.
\label{eq:kappa_thetabar_n}
\end{align}

The enthalpy-weighted mean square fluid 4-velocity around the bubble is defined by
\begin{equation}
\bar{U}_f^2 = \frac{\Delta e_Q}{w_n}.
\label{eq:ubarf2}
\end{equation}
The adiabatic index is defined as
\begin{equation}
\gamma \equiv \frac{c_P}{c_V},
\end{equation}
where $c_P$ is the specific heat capacity at constant pressure,
and $c_V$ is the specific heat capacity at constant volume.
For a classical monoatomic fluid $\gamma = \frac{5}{3}$,
and for an ultrarelativistic fluid $\gamma = \frac{4}{3}$.
The enthalpy-weighted mean square fluid 4-velocity around the bubble and
the kinetic energy density fraction of eq.~\eqref{eq:kinetic_energy_fraction} are linked by
\begin{equation}
K = \Gamma \bar{U}_f^2,
\label{eq:kinetic_energy_fraction2}
\end{equation}
where
\begin{equation}
\Gamma \equiv \frac{\bar{w}}{\bar{e}}.
\label{eq:mean_adiabatic_index}
\end{equation}
In the case of the ultrarelativistic bag model $\Gamma$ is the mean adiabatic index,
but for a more general model this may not be the case.

\chapter{Gravitational waves}
\label{ch:gw}
In this thesis the gravitational wave spectrum from the first order phase transition is calculated using the Sound Shell Model.
The Sound Shell Model is the process
that converts the fluid velocity and enthalpy profiles of chapter~\ref{ch:pt} to the gravitational wave spectra.
This process involves multiple steps.
First in section~\ref{gw_production} we investigate how the gravitational wave spectrum is produced by the shear stress.
Then in section~\ref{shear_stress} we investigate how the shear stress is produced by the sound waves.
We combine these results in section~\ref{gw_from_sound_waves}
to see how the gravitational wave spectrum is produced by sound waves.
Then in section~\ref{velocity_field} we convert the fluid velocity profiles of chapter~\ref{ch:pt} to sound waves.
Combining these, we have the gravitational wave spectra.
These four sections follow the contents and structure in the article~\cite{hindmarsh_gw_pt_2019} that introduces the Sound Shell Model.
In section~\ref{correction_factors} we extend the process beyond the base Sound Shell Model of the previous sections.
Then in section~\ref{omgw0} we convert the results to the frequencies today.
Finally, in section~\ref{noise} we define the LISA instrument noise
to compare the signal-to-noise ratios of the gravitational wave spectra to the LISA instrument noise.

\section{Gravitational wave production from shear stress}
\label{gw_production}
In this section we derive how the gravitational wave spectrum $\mathcal{P}_\text{gw}$ is produced by the shear stress correlator $U_\pi$.
First of all we assume that the time scale of the phase transition is much shorter than the Hubble time.
Therefore, we can neglect the expansion of the universe and assume that the background is Minkowski spacetime.
The fluid and the scalar field are sources of metric perturbations for the spacetime.
In the synchronous gauge these produce a change in the space-time interval
\cite[p. 7]{hindmarsh_gw_pt_2019}
\begin{equation}
ds^2 = -dt^2 + (\delta_{ij} + h_{ij}) dx^i dx^j.
\end{equation}
The metric perturbations $h_{ij}$ are sourced by shear stress,
which is the transverse-traceless part of the energy-momentum tensor $\Pi_{ij}$ of eq.~\eqref{eq:ep_tensor_general_matrix}.
This results in a wave equation for $h_{ij}$ with a source term,
\cites[eq. 3.1]{hindmarsh_gw_pt_2019}[eq. 1.24]{maggiore_gw_2008}
\begin{equation}
\ddot{h}_{ij} - \nabla^2 h_{ij} = 16 \pi G \Pi_{ij}.
\end{equation}
The energy-momentum tensor $T_{ij}$ consists of contributions from the fluid and the scalar field~\cite[p. 7]{hindmarsh_gw_pt_2019}
\begin{align}
T^f_{ij}    &= (e+p) \gamma^2 v_i v_j + p \delta_{ij}, \\
T^\phi_{ij} &= \delta_i \phi \delta_j \phi - \frac{1}{2}(\delta \phi)^2 \delta_{ij}.
\end{align}

Let us introduce the tensor
\cite[eq. 1.35]{maggiore_gw_2008}
\begin{equation}
P_{ij}(\bm{k}) \equiv \delta_{ij} - \hat{k}_i \hat{k}_j.
\end{equation}
This tensor is symmetric, transverse ($\hat{k}_i P_{ij}(\bm{k}) = 0$) and a projector ($P_{ik}P_{kj} = P_{ij}$), and its trace $P_{ii} = 2$.
Using $P_{ij}$ we can construct another tensor that we call the $\Lambda$ tensor
\cite[eq. 1.36]{maggiore_gw_2008}
\begin{equation}
\Lambda_{ij,kl}(\bm{k}) \equiv P_{ik}(\bm{k}) P_{jl}(\bm{k}) - \frac{1}{2} P_{ij}(\bm{k}) P_{kl}(\bm{k}).
\label{eq:Lambda}
\end{equation}
For further details on the $\Lambda$ tensor, please see~\cite[ch. 1.2]{maggiore_gw_2008}.

The spectral density of the time derivative of the perturbations, $P_{\dot{h}}(\bm{k},t)$, is defined by
\cite[eq. 3.4]{hindmarsh_gw_pt_2019}%
\footnote{In~\cite[eq. 3.4]{hindmarsh_gw_pt_2019} the second $\mathbf{k}$ is a typo and should be $\mathbf{k}'$.}
\begin{equation}
\langle \dot{h}_{\bm{k}}^{ij}(t) \dot{h}_{\bm{k}'}^{ij}(t) \rangle = P_{\dot{h}}(\bm{k},t) (2\pi)^3 \delta (\bm{k} + \bm{k}').
\label{eq:hbracket}
\end{equation}

The energy density of gravitational waves is given by
\cites[eq. 3.3]{hindmarsh_gw_pt_2019}[eq. 1.135, 7.193]{maggiore_gw_2008}
\begin{align}
e_{gw}
&= \frac{1}{32 \pi G} \int dk^3 \langle \dot{h}_{ij} \dot{h}_{ij} \rangle \\
&= \frac{1}{32 \pi G} \int dk \frac{k^2}{2 \pi^2} P_{\dot{h}}(k).
\end{align}
In many textbooks and articles the integral in the first expression is implicit.
The factor of $\frac{k^2}{2\pi^2}$ results from the conversion to spherical coordinates.
The critical energy density corresponding to a flat universe is given by
\cite[eq. 7.196]{maggiore_gw_2008}
\begin{equation}
e_c = \frac{3 H^2}{8 \pi G}.
\label{eq:e_crit}
\end{equation}
Let us define energy density of gravitational waves relative to the critical energy density.
This is the dimensionless quantity
\begin{equation}
\Omega_{gw} \equiv \frac{e_{gw}}{e_c}.
\label{eq:omega_gw}
\end{equation}

The gravitational wave power spectrum is defined by
\cite[eq. 3.45]{hindmarsh_gw_pt_2019}
\begin{equation}
\mathcal{P}_{gw}(k)
\equiv \frac{d \Omega_{gw}}{d \ln (k)}
= \frac{1}{12 H^2} \frac{k^3}{2\pi} P_{\dot{h}}(k)
= \frac{1}{12 H^2} \mathcal{P}_{\dot{h}}(k),
\label{eq:gw_pow_spec}
\end{equation}
where the first step is a direct result from the equations above.
It should be noted that in~\cite[eq. 3.6, eq. 3.46]{hindmarsh_gw_pt_2019} it is presumed that $\bar{\rho}=e_c$.
A spectral density can be converted to a power spectrum with
\cite[eq. 4.18]{hindmarsh_gw_pt_2019}
\begin{equation}
\mathcal{P}(k) = \textcolor{gray}{(2)} \frac{k^3}{2 \pi^2} P(k).
\label{eq:pow_spec}
\end{equation}
The factor of two is due to the fact that the velocity Fourier transform includes waves moving in both directions.%
\footnote{See also~\cite[p. 338]{maggiore_gw_2008}.}
Whether it is included or not is dependent on the context.
Therefore, the gravitational wave power spectrum can be obtained from the spectral density $P_{\dot{h}}$ as
\begin{equation}
\mathcal{P}_{gw}(k) = \frac{1}{12 H^2} \frac{k^3}{2\pi^2} P_{\dot{h}}.
\label{eq:gw_pow_spec2}
\end{equation}
To obtain the gravitational wave spectrum $\mathcal{P}_{gw}(k)$,
we therefore need to obtain an expression for $P_{\dot{h}}$ that we can compute from the fluid shell profile.
For this we will need quite a few similar intermediate quantities.
To avoid confusion, their symbols and names are listed in table~\ref{table:symbols}.

\begin{table}[ht]
\caption{Symbols and explanations of the spectral quantities}
\begin{tabular}{l|l|l}
Symbol & Explanation & Eq.\\
\hline
$P_{\dot{h}}$ & Spectral density of $\dot{h}$ & \eqref{eq:p_dot_h} \\
$P_v$ & Spec. den. of the plane wave components of the velocity field & \eqref{eq:spec_den_v} \\
$\tilde{P}_v$ & Scaled velocity spectral density & \eqref{eq:tilde_p_v} \\
$\mathcal{P}_{\tilde{v}}$ & Velocity power spectrum & \eqref{eq:pow_v} \\
$\tilde{P}_\text{gw}$ & Spectral density of the gravitational wave power spectrum & \eqref{eq:spectral_density} \\
$\mathcal{P}_\text{gw}$ & Gravitational wave power spectrum & \eqref{eq:gw_pow_spec2} \\
$\dot{P}_{\dot{h}}$ & Growth rate of the spectral density of $\dot{h}$ & \eqref{eq:dot_p_dot_h} \\
$\mathcal{P}'_{\text{gw}}$ & Growth rate of the GW spectrum relative to the Hubble rate & \eqref{eq:pow_gw_prime}
\end{tabular}
\label{table:symbols}
\end{table}

There exists a solution to the gravitational wave equation in the form of
\cite[eq. 3.2]{hindmarsh_gw_pt_2019}
\begin{equation}
h_{ij} (\bm{k},t) = (16 \pi G) \Lambda_{ij,kl}(\bm{k}) \int_0^t dt' \frac{\sin [k(t-t')]}{k} T_{kl}(\bm{k},t').
\label{eq:h_ij}
\end{equation}
Since we are operating in the transverse-traceless gauge, it is sufficient to replace $T_{ij}$ with the tensor
\cite[eq. 3.7]{hindmarsh_gw_pt_2019}
\begin{equation}
\tau_{ij} \equiv \gamma^2 w v_i v_j + \partial_ \phi \partial_j \phi.
\label{eq:tau_ij}
\end{equation}
This will be approximated in eq.~\eqref{eq:tau_ij_approx}.
Using eq.~\eqref{eq:h_ij} and~\eqref{eq:tau_ij} we have
\cite[eq. 3.8]{hindmarsh_gw_pt_2019}
\begin{multline}
\langle \dot{h}_{\bm{k}_1}^{ij}(t) \dot{h}_{\bm{k}_2}^{ij}(t) \rangle \\
= (16 \pi G)^2 \int_0^t dt_1 dt_2 \cos [k_1(t-t_1)] \cos [k_2(t-t_2)] \Lambda_{ij,kl}(\bm{k})
\langle \tau_{\bm{k}_1}^{ij}(t_1) \tau_{\bm{k}_2}^{kl}(t_2) \rangle.
\end{multline}
Let us define the unequal time correlator (UETC) of the fluid shear stress $U_\Pi$ with
\cite[eq. 3.9]{hindmarsh_gw_pt_2019}
\begin{equation}
\Lambda_{ij,kl}(\bm{k}) \langle \tau_{\bm{k}_1}^{ij}(t_1) \tau_{\bm{k}_2}^{kl}(t_2) \rangle
= U_\Pi (k_1, t_1, t_2) (2 \pi)^3 \delta(\bm{k}_1 + \bm{k}_2).
\label{eq:hbracket2}
\end{equation}
Using this we can make simple substitutions to eq.~\eqref{eq:hbracket} and~\eqref{eq:hbracket2},
resulting in an expression for the spectral density as
\cite[eq. 3.10]{hindmarsh_gw_pt_2019}
\begin{equation}
P_{\dot{h}} (k,t) = (16 \pi G)^2 \int_0^t dt_1 \int_0^t dt_2 \cos [k(t-t_1)] \cos [k(t-t_2)] U_\Pi (k, t_1, t_2).
\label{eq:p_dot_h}
\end{equation}
Averaging over oscillations at wavenumber $k$,
\begin{equation}
P_{\dot{h}} (k,t) = (16 \pi G)^2 \frac{1}{2} \int_0^t dt_1 \int_0^t dt_2 \cos \left( k(t_1 - t_2) \right) U_\Pi (k, t_1, t_2).
\label{eq:p_dot_h_avg}
\end{equation}
This way we can compute the gravitational wave spectrum $\mathcal{P}_\text{gw}$ from the fluid shear stress UETC $U_\Pi$.

\section{Shear stress from sound waves}
\label{shear_stress}
In this section we investigate how the unequal time correlator for the shear stress (UETC) can be computed from the sound waves in the fluid.
Let us start by defining the fluid stress-energy tensor $\tau_{ij}$.
The $\partial_i \phi \partial_j \phi$ term of eq.~\eqref{eq:tau_ij} is non-zero only at the phase bondary,
which is thin and therefore negligible compared to the fluid shell.
We also approximate that once the bubbles have collided and merged,
the enthalpy is constant throughout the fluid shell: $w = e + p \approx \bar{w}$,
and that the fluid velocities are non-relativistic, resulting in $\gamma(v) = 1$.
Therefore eq.~\eqref{eq:tau_ij} approximates as
\cite[eq. 3.12]{hindmarsh_gw_pt_2019}
\begin{equation}
\tau_{ij} \approx \bar{w} v_i v_j.
\label{eq:tau_ij_approx}
\end{equation}

Let us define a couple of mathematical tools and quantities.
The difference of the wavevectors $\mathbf{q}$ and $\mathbf{k}$,
\begin{equation}
\tilde{\mathbf{q}} = \mathbf{q} - \mathbf{k}.
\label{eq:tilde_q}
\end{equation}
Using this their product can be expressed as
\begin{equation}
\mu \equiv \hat{\mathbf{q}} \cdot \hat{\mathbf{k}} = \frac{q^2 + k^2 - \tilde{q}^2}{2kq}.
\label{eq:mu_ssm}
\end{equation}
The length of a vector is denoted as
\begin{equation}
q \equiv |\mathbf{q}|.
\end{equation}
The unit vectors are defined as
\begin{equation}
\hat{\mathbf{q}} \equiv \frac{\mathbf{q}}{|\mathbf{q}|} \Rightarrow
\hat{q}^i \equiv \frac{q^i}{q}.
\end{equation}
The angular frequencies $\omega$ and $\tilde{\omega}$ are defined as
\cite[p. 10]{hindmarsh_gw_pt_2019}
\begin{equation}
\omega \equiv c_s q, \quad \tilde{\omega} \equiv c_s \tilde{q}.
\end{equation}
The $\Lambda$ tensor of eq.~\eqref{eq:Lambda} has the property that
\cite[eq. 3.17]{hindmarsh_gw_pt_2019}
\begin{equation}
\Lambda_{ij,kl}(\mathbf{k}) \hat{q}^i \hat{\tilde{q}}^j \hat{\tilde{q}}^k \hat{q}^l = \frac{1}{2}(1 - \mu^2)^2 \frac{q^2}{\tilde{q}^2}.
\end{equation}
For later use let us also define
\cite[p. 11]{hindmarsh_gw_pt_2019}
\begin{equation}
t_+ \equiv \frac{t_1 + t_2}{2}, \quad t_- \equiv t_1 - t_2.
\label{eq:t_plus_minus}
\end{equation}

Using these tools Hindmarsh et al. have derived in~\cite{hindmarsh_gw_pt_2019} that the UETC can be given as
\cite[eq. 3.32]{hindmarsh_gw_pt_2019}
\begin{equation}
U_\Pi(k, t_1, t_2) = 4 \bar{w}^2 \int \frac{d^3 q}{(2\pi)^3} \frac{q^2}{\tilde{q}^2} (1 - \mu^2) P_v(q) P_v(\tilde{q}) \cos (\omega t_-) \cos (\tilde{\omega} t_-).
\end{equation}
Further by performing an angular integration and changing the integration variable using eq.~\eqref{eq:mu_ssm}, we have
\cite[eq. 3.34]{hindmarsh_gw_pt_2019}
\begin{equation}
U_\Pi (k, t_1, t_2) = \frac{4 \bar{w}^2}{4 \pi^2 k} \int_0^\infty \int_{|q-k|}^{q+k} d\tilde{q} q \tilde{q}
\frac{q^2}{\tilde{q}^2} (1-\mu^2)^2
P_v(q) P_v(\tilde{q})
\cos (\omega t_-) \cos (\tilde{\omega} t_-).
\label{eq:uetc_final}
\end{equation}
The contents of $P_v(q)$ will be defined later in eq.~\eqref{eq:spec_den_v}.

\section{Gravitational wave power spectrum from sound waves}
\label{gw_from_sound_waves}
In this section we derive the gravitational wave spectrum from the unequal time correlator (UETC).
The UETC of eq.~\eqref{eq:uetc_final} can be substituted to~\eqref{eq:p_dot_h_avg},
resulting in
\cite[eq. 3.35]{hindmarsh_gw_pt_2019}
\begin{equation}
P_{\dot{h}} (k,t) = (16 \pi G)^2 \frac{4 \bar{w}}{4\pi^2 k}
\int_0^\infty dq \int_{|q-k|}^{q+k} d\tilde{q} q \tilde{q} \frac{q^2}{\tilde{q}^2} (1-\mu^2)^2
P_v(q) P_v(\tilde{q}) \Delta(t,k,q,\tilde{q}),
\end{equation}
where
\begin{equation}
\Delta(t,k,q,\tilde{q}) \approx \frac{1}{2} \int_0^t dt_1 \int_0^t dt_2 \cos(kt_-)\cos(\omega t_-)\cos(\tilde{\omega}t_-).
\end{equation}
The approximation comes from averaging over the number of oscillations at wavenumber $k$ in eq.~\eqref{eq:p_dot_h_avg},
and $t_-$ is defined in eq.~\eqref{eq:t_plus_minus}.
Using eq.~\eqref{eq:t_plus_minus} one can also change the integration variable to $t_0$, resulting in
\begin{equation}
\Delta(t,k,q,\tilde{q}) = \frac{1}{2} \int_0^t dt_+ \int_{-2t_+}^{2t_+} dt_- \cos(kt_-) \cos(\omega t_-) \cos(\tilde{\omega} t_-).
\end{equation}
Therefore its growth rate is
\begin{equation}
\dot{\Delta}(t,k,q,\tilde{q}) \equiv \frac{d}{dt} \Delta(t,k,q,\tilde{q}) = \frac{1}{2} \int_{-2t}^{2t} dt_- \cos(kt_-) \cos(\omega t_-) \cos(\tilde{\omega} t_-).
\label{eq:delta_dot}
\end{equation}
Using a trigonometric identity for the product of cosines,
at large $t$, eq.~\eqref{eq:delta_dot} asymptotes to a $\delta$-function as
\begin{equation}
\lim_{t\rightarrow\infty} \dot{\Delta}(t,k,q,\tilde{q}) = \frac{\pi}{8} \Sigma_{\pm\pm\pm} \delta(\pm k \pm \omega \pm \tilde{\omega}).
\end{equation}
Of these combinations, only $k - \omega - \tilde{\omega}$ and $-(k - \omega - \tilde{\omega})$ can vanish,
and therefore
\begin{equation}
\lim_{t\rightarrow\infty} \dot{\Delta}(t,k,q,\tilde{q}) = \frac{\pi}{4} \delta (k - \omega - \tilde{\omega}).
\label{eq:delta_dot_lim}
\end{equation}
Beyond these approximations, these two terms are dominant,
but the other terms do have a finite contribution as well~\cites{sharma_shallow_2023}{pol_characterization_2023}.
Inserting this to eq.~\eqref{eq:p_dot_h_avg} gives us the asymptotic growth rate of the spectral density
\begin{equation}
\lim_{t \rightarrow \infty} \dot{P}_{\dot{h}}(k,t)
= (16 \pi G)^2 \frac{4 \bar{w}^2}{4 \pi^2 k} \int_0^\infty dq \int_{|q-k|}^{q+k} d\tilde{q} q \tilde{q} \frac{q^2}{\tilde{q}^2} (1 - \mu^2)^2 P_v(q) P_v(\tilde{q}) \frac{\pi}{4} \delta (k - \omega - \tilde{\omega}).
\end{equation}
The $\delta$-function of eq.~\eqref{eq:delta_dot_lim} has restricted us to
\begin{equation}
k - \omega - \tilde{\omega} = k - c_s (q - \tilde{q}) = 0,
\end{equation}
and therefore we have
\begin{equation}
\tilde{q} = \frac{k}{c_s} - q.
\label{eq:tilde_q2}
\end{equation}
Using this we can define
\begin{equation}
q_\pm \equiv \frac{k(1 \pm c_s)}{2 c_s}.
\end{equation}
Inserting eq.~\eqref{eq:tilde_q2} to~\eqref{eq:mu_ssm} gives
\begin{equation}
\mu = \frac{2qc_s - k(1 - c_s^2)}{2qc_s^2}.
\end{equation}
We can now integrate over $q$. Since the expression is no longer time-dependent,
we can denote it as
\begin{equation}
\dot{P}_{\dot{h}}(k) = (16 \pi G)^2 \frac{\bar{w}}{4 \pi k c_s} \int_{q_-}^{q_+} \frac{q^3}{\tilde{q}} (1 - \mu^2)^2 P_v(q) P_v(\tilde{q}).
\label{eq:dot_p_dot_h}
\end{equation}

We can now define the scaled velocity spectral density $\tilde{P}_v$ as
\begin{equation}
\tilde{P}_v (qL_f) \equiv \frac{P_v(q)}{L_f^3 \bar{U}_f^2},
\label{eq:tilde_p_v}
\end{equation}
where $L_f$ is a length scale in the velocity field, and $\bar{U}_f$ is the RMS fluid velocity.
Let us also define the additional quantities
\begin{equation}
y = kL_f, \quad z = qL_f, \quad z_\pm = y \frac{1 \pm c_s}{2 c_s}.
\label{eq:gw_yz}
\end{equation}
Therefore, the asymptotic growth rate of the spectral density can be written as
\begin{equation}
\dot{P}_{\dot{h}}(y) =
\left( 16 \pi G \bar{w} \bar{U}_f^2 \right)^2
\frac{L_f^4}{4 \pi y c_s}
\int_{z_-}^{z_+} dz
\frac{z^3}{\frac{y}{c_s} - z}
(1 - \mu^2)^2
\tilde{P}_v (z) \tilde{P}_v \left( \frac{y}{c_s} - z \right).
\end{equation}
The growth rate of the gravitational wave spectrum of eq.~\eqref{eq:gw_pow_spec2} relative to the Hubble rate is
\cite[eq. 3.46]{hindmarsh_gw_pt_2019}
\begin{equation}
\mathcal{P}'_{\text{gw}} \equiv \frac{1}{H} \frac{d}{dt} \mathcal{P}_{\text{gw}}
= 3 \left( \Gamma \bar{U}_f^2 \right)^2 (HL_f) \frac{(kL_f)^3}{2 \pi^2} \tilde{P}_{\text{gw}} (kL_f),
\label{eq:pow_gw_prime}
\end{equation}
where we have used the $\Gamma$ of eq.~\eqref{eq:mean_adiabatic_index} and $e_c$ of eq. eq.~\eqref{eq:e_crit} and approximated that $\bar{e} = e_c$.
We have also used the $y$ and $z$ of eq.~\eqref{eq:gw_yz}, and using these we have defined $\tilde{P}_{\text{gw}}$,
the dimensionless spectral density function
\cite[eq. 3.47]{hindmarsh_gw_pt_2019}%
\footnote{In~\cite{hindmarsh_gw_pt_2019} the $\bar{P}_v$ is a typo and should be $\tilde{P}_v$.}
\begin{equation}
\tilde{P}_\text{gw} (y) \equiv \frac{1}{4\pi yc_s} \left(\frac{1-c_s^2}{c_s^2}\right)^2
\int_{z_-}^{z_+} \frac{dz}{z}
\frac{(z-z_+)^2(z-z_-)^2}{z_+ + z_- - z}
\tilde{P}_v (z) \tilde{P}_v (z_+ + z_- z).
\label{eq:spectral_density}
\end{equation}
The total gravitational wave power spectrum for a stationary velocity power spectrum with a lifetime $\tau_v = t - t_0$ is given by
\cite[eq. 3.48]{hindmarsh_gw_pt_2019}%
\footnote{In~\cite{hindmarsh_gw_pt_2019} the $\tilde{P}_{GW}$ is a typo and should be $\tilde{P}_\text{gw}$.}
\begin{equation}
\Omega_\text{gw}^\text{ssm}
\equiv \mathcal{P}_\text{gw}(k)
= \int_{t_0}^{t} dt H \mathcal{P}'_\text{gw}
= 3 \left( \Gamma \bar{U}_f^2 \right)^2 (H \tau_v)(H L_f) \frac{(kL_f)^3}{2\pi^2} \tilde{P}_\text{gw} (kL_f),
\label{eq:gw_pow_spec3}
\end{equation}
where $\Gamma$ is the mean adiabatic index of eq.~\eqref{eq:mean_adiabatic_index}.
This is the equation used by PTtools to compute the gravitational wave spectrum.
The choice of $c_s$ in eq.~\eqref{eq:spectral_density} is important and non-trivial,
since beyond the bag model $c_s = c_s(T(w(\xi),\phi(\xi)))$.
Therefore, for an enthalpy-dependent sound speed the formula is not exact.
We approximate $c_s$ here as $c_s(T(\bar{w},\phi_b),\phi_b)$,
as $\bar{w}$, the mean enthalpy after the phase transition of eq.~\eqref{eq:wbar},
gives a reasonable estimate on the temperature of the fluid.
For the constant sound speed model of section~\ref{const_cs} this gives simply $c_{s,b}$.

\section{Velocity field from superposition of single-bubble fluid shells}
\label{velocity_field}
Now in the equations~\eqref{eq:tilde_p_v},~\eqref{eq:spectral_density} and~\eqref{eq:gw_pow_spec3}
we have a system that takes in the velocity spectral density $P_v$ and outputs the gravitational wave spectrum.
Therefore, the missing piece is to compute the velocity spectral density $P_v$ from the fluid shell profile.
The Sound Shell Model does not account for the collision dynamics,
but approximates that the bubble completely disappears when half of it has merged with another advancing region of the stable phase.
The velocity field arises from the superposition of fields produced by $N_b$ individual bubbles,
\cite[eq. 4.1]{hindmarsh_gw_pt_2019}
\begin{equation}
v_i(\mathbf{x},t) = \sum_{n=1}^{N_b} v_i^{(n)} (\mathbf{x},t).
\end{equation}
The dimensionless coordinate $\xi$ is defined as in eq.~\eqref{eq:xi},
\begin{equation}
\xi \equiv \frac{R^{(n)}}{T^{(n)}},
\end{equation}
where $T^{(n)}$ is the time since the nucleation of the $n$th bubble, given by
\begin{equation}
T^{(n)} = t - t^{(n)}.
\label{eq:bubble_lifetime}
\end{equation}
The velocity field of each bubble is radial with respect to its center $\mathbf{x}_n$, it can be writen as
\cite[eq. 4.2]{hindmarsh_gw_pt_2019}
\begin{equation}
v_i^{(n)}(\mathbf{x},t) = \frac{R_i^{(n)}}{R^{(n)} v_\text{ip}(\xi)},
\end{equation}
where $R_i^{(n)} = x_i - x_i^{(n)}$ is the distance of the bubble surface from its centre.
The Fourier transform of the velocity field of a single bubble is given by
\cite[eq. 4.3]{hindmarsh_gw_pt_2019}
\begin{align}
\tilde{v}_i^{(n)} (\mathbf{q},t)
&= \int d^3 x v_i^{(n)} (\mathbf{x},t) e^{-i \mathbf{q} \cdot \mathbf{x}} \\
&= \int d^3 R^{(n)} \frac{R_i^{(n)}}{R^{(n)}} v_\text{ip}(\xi) e^{-i \mathbf{q} \cdot \mathbf{R}^{(n)}}.
\end{align}
Let us define the dimensionless wavenumber $z$ as
\begin{equation}
z^i = q^i T^{(n)}.
\end{equation}
By changing the integration variable from $R^{(n)}$ to $\xi$ we get
\cite[eq. 4.4, eq. 4.6]{hindmarsh_gw_pt_2019}
\begin{align}
\tilde{v}_i^{(n)} (\mathbf{q},t)
&= e^{-i \mathbf{q} \cdot x^{(n)}} i (T^{(n)})^3 \frac{\partial}{\partial z_i} \left(
\int d^3 \xi \frac{1}{\xi} v_\text{ip}(\xi) e^{-i z^i \xi^i} \right) \\
&= e^{-i \mathbf{q} \cdot x^{(n)}} i (T^{(n)})^3 \hat{z}^i f'(z),
\end{align}
where we have denoted the part in the parentheses as $f(z)$, and $f'(z) \equiv \frac{d}{dz}f(z)$.
Performing the angular integration in $f(z)$ results in
\cite[eq. 4.5]{hindmarsh_gw_pt_2019}
\begin{equation}
f(z) \equiv \int d^3\xi \frac{1}{\xi} v_\text{ip}(\xi) e^{-iz^i \xi^i}
= \frac{4\pi}{z} \int_0^\infty d\xi v_\text{ip}(\xi) \sin(z\xi).
\label{eq:ssm_f}
\end{equation}
Performing similar steps for the energy perturbation variable $\lambda$ of eq.~\eqref{eq:lambda},
we get its Fourier transformation
\begin{equation}
\tilde{\lambda}^{(n)}(\mathbf{q},t) = e^{-i \mathbf{q} \cdot \mathbf{x}} (T^{(n)})^3 l(z),
\end{equation}
where similar to $f(z)$ we have $l(z)$ given by
\cite[eq. 4.8]{hindmarsh_gw_pt_2019}
\begin{align}
l(z) \equiv \frac{4 \pi}{z} \int_0^\infty d\xi \lambda_\text{ip}(\xi) \sin(z\xi),
\label{eq:ssm_l}
\end{align}
where $\lambda_\text{ip}$ is the energy fluctuation variable $\lambda$ of eq.~\eqref{eq:lambda} for a single bubble.
The integrations over the sines in these functions are the sine transform of the Sound Shell Model.
Assuming that the entire fluid perturbation characterised by $f(z)$ and $l(z)$ becomes the initial condition for a sound wave at the collision time $t_i^{(n)}$,
its contribution to the plane wave is
\cite[eq. 4.9]{hindmarsh_gw_pt_2019}
\begin{equation}
v_{\mathbf{q},i}^{(n)} = i (T_i^{(n)})^3 \hat{z}_i e^{i \omega t_i - i \mathbf{q} \cdot \mathbf{x}^{(n)}} A(z).
\end{equation}
These can be combined to the wavefunction launched by a single bubble,
\cite[eq. 4.10]{hindmarsh_gw_pt_2019}
\begin{equation}
A(z) \equiv \frac{1}{2} \left( f'(z) + i c_s l(z) \right).
\end{equation}
Since $f'$ and $\lambda$ are real,
\cite[eq. 4.11]{hindmarsh_gw_pt_2019}
\begin{equation}
|A(z)|^2 = \frac{1}{4} \left( f'(z)^2 + (c_s l(z))^2 \right).
\end{equation}

Now that we have the plane wave created by a single bubble,
we can combine these to get the overall velocity power spectrum.
The plane wave correlation function for $N_b$ randomly-placed bubbles in a volume $\mathcal{V}$ is given by
\cite[eq. 4.12]{hindmarsh_gw_pt_2019}
\begin{equation}
\langle v_{\mathbf{q}_1}^i v_{\mathbf{q}_2}^{*j} \rangle
= \sum_{m=1}^{N_b} \sum_{n=1}^{N_b} \langle
(T_i^{(m)})^3 (T_i^{(n)})^3 \hat{z}^i \hat{z}'^{j} A(z) A^*(z')
e^{-i \mathbf{q_1} \cdot \mathbf{x}^{(m)} + i \mathbf{q}_2 \cdot \mathbf{x}^{(n)}}
e^{i (\omega_1 - \omega_2) t_i}
\rangle.
\end{equation}
The averaging $\langle \rangle$ is over the ensemble of bubble locations $\mathbf{x}^{(n)}$, nucleation times $t^{(n)}$ and collision times $t_i^{(n)}$.

First averaging over bubbles nucleated between $t'$ and $t' + dt'$, and colliding between $t_i$ and $t_i + dt_i$, we have
\cite[eq. 4.13]{hindmarsh_gw_pt_2019}
\begin{equation}
\sum_{m=1}^{N_b} \sum_{n=1}^{N_b} \langle e^{-i \mathbf{q}_1 \cdot \mathbf{x}^{(m)} + i \mathbf{q}_2 \cdot \mathbf{x}^{(n)}} \rangle
= d^2 P \frac{N_b}{\mathcal{V}} (2\pi)^3 \delta(\mathbf{q}_1 \mathbf{q}_2),
\label{eq:bubble_average}
\end{equation}
where $d^2 P(t', t_i)$ is the joint probability for nucleating and colliding in the given time ranges.
The $\delta$-function results in $\omega_1 = \omega_2$, which removes the $e^{i(\omega_1 - \omega_2)t_i}$ term.
Therefore, the result is not dependent on the absolute collision time $t_i$,
but only on the average over the bubble lifetimes $T_i$.
Let us denote the probability density distribution of lifetimes as
\begin{equation}
n(T_i) \equiv \frac{N_b}{\mathcal{V}} \frac{dP(T_i)}{dT_i}.
\end{equation}
Using these we can rewrite eq.~\eqref{eq:bubble_average} as
\begin{equation}
\langle v_{\mathbf{q}_1}^i v_{\mathbf{q}_2}^{*j} \rangle = \int dT_i n(T_i) T_i^6 \hat{z}^i \hat{z}^j |A(z)|^2 (2\pi)^2 \delta(\mathbf{q}_1 - \mathbf{q}_2).
\end{equation}
Let us define the mean bubble separation $R_*$, for which
\begin{equation}
\lim_{\mathcal{V}\rightarrow\infty} R_*^3 = \frac{N_b}{\mathcal{V}}.
\end{equation}
Therefore,
\begin{equation}
\int n(T_i) dT_i = \frac{1}{R_*^3}.
\end{equation}
These result in
\cite[eq. 4.15]{hindmarsh_gw_pt_2019}
\begin{equation}
n(T_i) dT_i = \frac{\beta}{R_*^3} \nu(\beta T_i) dT_i,
\end{equation}
where $\nu (\beta T)$ is the bubble lifetime distribution function, which is normalised so that $\int \nu(x) dx = 1$.
The rate $\beta$ is the nucleation rate parameter, for which
\cite[eq. 4.16]{hindmarsh_gw_pt_2019}
\begin{equation}
\beta = (8 \pi)^{\frac{1}{3}} \frac{v_\text{wall}}{R_*}.
\end{equation}
This is the definition of $\beta$ in the case of simultaneous nucleation.
For the case of exponential nucleation and further information on the nucleation rate in general,
please see~\cite[section 4.2]{hindmarsh_gw_pt_2019}.
Similarly as in eq.~\eqref{eq:hbracket}, we can now derive the spectral density of the plane wave components of the velocity field as
\cite[eq. 4.17]{hindmarsh_gw_pt_2019}
\begin{equation}
P_v(q) = \frac{1}{\beta^6}{R_*^3} \int d\tilde{T} \nu(\tilde{T}) \tilde{T}^6 |A(\frac{\tilde{T}q}{\beta})|^2,
\label{eq:spec_den_v}
\end{equation}
where $\tilde{T} \equiv \beta T_i$.
Using eq.~\eqref{eq:pow_spec} we can convert this to the velocity power spectrum
\cite[eq. 4.18]{hindmarsh_gw_pt_2019}
\begin{equation}
\mathcal{P}_{\tilde{v}}(q)
= 2 \frac{q^3}{2\pi^2} P_v(q)
= \frac{2}{(\beta R_*)^3} \frac{1}{2\pi^2} \left(\frac{1}{\beta}\right)^3
\int d \tilde{T} \eta(\tilde{T}) \tilde{T}^6 \left| A \left( \frac{\tilde{T} q}{\beta} \right) \right|^2.
\label{eq:pow_v}
\end{equation}
This can be inserted to eq.~\eqref{eq:gw_pow_spec3} to get the gravitational wave spectrum.
Now we have the full Sound Shell Model to convert from the fluid velocity profile to the gravitational wave spectrum.

\section{Correction factors}
\label{correction_factors}
The Sound Shell Model is an approximation, although a highly useful one.
To improve upon its results, several extensions can be applied.
Giombi et al. have investigated the effects of changing the sound speed on the gravitational wave power spectrum using analytical calculations.
To implement this extension, we need the barotropic equation of state parameter
\cite[p. 3]{giombi_cs_2024}
\begin{equation}
\omega(T,\phi) \equiv \frac{p(T,\phi)}{e(T,\phi)}.
\end{equation}
Giombi et al. define the quantity $\nu$, which we shall call as $\nu_\text{gdh2024}$, as
\cite[eq. 2.11]{giombi_cs_2024}
\begin{equation}
\nu_\text{gdh2024} \equiv \frac{1 - 3\omega}{1 + 3\omega}.
\label{eq:nu_gdh2024}
\end{equation}
Giombi et al. use $\frac{\Delta \eta_v}{\eta_*}$ as the source duration,
whereas the flow lifetime $\tau_v$ is a more common quantity.
It is proportional to known quantities as
\cites[p. 3]{hindmarsh_gw_pt_2019}[p. 6]{gowling_lisa_2021}
\begin{equation}
\tau_v \sim \frac{R_*}{\bar{U}_f} \sim \frac{R_*}{\sqrt{K}},
\end{equation}
where $R_*$ is the mean bubble separation, and
$K$ is the kinetic energy fraction of eq.~\eqref{eq:kinetic_energy_fraction2},
which can be computed from the fluid shell.
The Hubble-scaled mean bubble spacing is defined as
\cite[eq. 2.2]{gowling_lisa_2021}
\begin{equation}
r_* \equiv H_n R_*.
\end{equation}
To convert between $r_*$ and the source duration by Giombi et al.,
we define the source lifetime multiplier $\lambda$ as
\begin{equation}
\frac{\Delta \eta}{\eta_*} = \lambda \frac{2 r_*}{\sqrt{K}}.
\end{equation}
We define the source lifetime factor $\Lambda$ as
\cite[eq. 3.13]{giombi_cs_2024}
\begin{equation}
\Lambda \equiv \frac{1}{1 + 2\nu} \left(1 - \left(1 + \frac{\Delta \eta}{\eta_*} \right) \right)^{-1-2\nu},
\end{equation}
where $\nu \equiv \nu_\text{gdh2024}$ of eq.~\eqref{eq:nu_gdh2024}.
With this we can describe the corrected spectral density by Giombi et al. simply as
\begin{equation}
\tilde{P}_\text{gw,corr.}(k)
\equiv \Lambda \tilde{P}_\text{gw}(k).
\end{equation}

Another thing that we need to take into account is that the Sound Shell Model is an approximation even for the bag model,
and to get more accurate results,
we need to apply a suppression factor $\Sigma(v_\text{w},\alpha_n)$ derived from comparing Sound Shell Model results to 3D hydrodynamic simulations,
\cite[eq. 2.9]{gowling_lisa_2021}
\begin{equation}
\Omega_\text{gw}(z) = \Omega_\text{gw}^\text{ssm}(z) \Sigma(v_\text{w},\alpha_n),
\end{equation}
where $\Omega_\text{gw}^\text{ssm}(z)$ is the Sound Shell Model prediction from eq.~\eqref{eq:gw_pow_spec3}.
Both the source lifetime factor $\Lambda$ and the suppression factor $\Sigma$ are implemented in PTtools,
and correspondingly in the results of chapter~\ref{ch:results}.

\section{Gravitational wave power spectra today}
\label{omgw0}
Inserting the result~\eqref{eq:pow_v} from the Sound Shell Model to eq.~\eqref{eq:gw_pow_spec3} gives the gravitational wave spectrum $\mathcal{P}_\text{gw}(k)$.
In this section we convert this to the gravitational wave spectrum today, $\Omega_{\text{gw},0}$, which is an observable quantity.
The power attenuation following the end of the radiation era is given by
\cites[eq. 2.11]{gowling_lisa_2021}[eq. 19]{caprini_detecting_2020}%
\footnote{In~\cite[eq. 2.11]{gowling_lisa_2021} the $\frac{4}{9}$ is a typo and should be $\frac{4}{3}$ as in~\cite[eq. 19]{caprini_detecting_2020}.}
\begin{equation}
F_{\text{gw},0} = \Omega_{\gamma,0} \left( \frac{g_{s0}}{g_{s*}} \right)^\frac{4}{3} \frac{g_*}{g_0}.
\end{equation}
Here the density parameter of photons today, $\Omega_{\gamma,0} \approx 1.0995 \cdot 10^{-4}$.
The degrees of freedom today are $g_0 = 2, g_{s0} \approx 3.91$.
In the Standard Model $g_{s*} = g_*$ for $T > 0.1 \ \text{MeV}$,
and in PTtools these can be extracted from the given equation of state or specified by the user.
The Hubble rate at the phase transition redshifted to today is given by
\cites[eq. 2.13]{gowling_lisa_2021}[eq. 31]{caprini_detecting_2020}
\begin{equation}
f_{*,0} = 2.6 \cdot 10^{-6} \text{Hz} \left( \frac{T_n}{100 \text{GeV}} \right) \left( \frac{g_*}{100} \right)^\frac{1}{6}.
\end{equation}
Using this we can convert from the dimensionless wavenumber $z$ to frequency today by taking into account the redshift
\cite[eq. 2.12]{gowling_lisa_2021}
\begin{equation}
f = \frac{z}{r_*} f_{*,0}.
\end{equation}
The power spectrum today at the physical frequency $f$ is
\cite[eq. 2.10]{gowling_lisa_2021}
\begin{equation}
\Omega_{\text{gw},0}(f) = F_{\text{gw},0} \Omega_\text{gw}(z(f)).
\end{equation}
Now we have defined the complete process from converting the fluid shell profile to the observable gravitational wave spectrum today.

\section{Instrument noise}
\label{noise}
To determine whether a phase transition can be resolved by LISA,
we need to compare its gravitational wave power spectrum to the power spectrum of the noise.
The LISA noise consists of the instrument noise and noise from the astrophysical foreground.
The most notable astrophysical noise sources are the extragalactic compact binaries and unresolved galactic compact binaries.
These have been left out from this thesis for simplicity, but are included in PTtools.
\cites{gowling_lisa_2021}{pttools}

The LISA instrument noise is expected to consist primarily of two main sources:
the test mass acceleration noise (acc), and the optical metrology noise (oms).
The optical metrology noise is caused by single link optical path-length fluctuations.
Their noise target for LISA is
\cites[eq. 3.2]{gowling_lisa_2021}[eq. 54]{smith_lisa_2019}
\begin{equation}
P_\text{oms} = \left( \frac{1.5 \cdot 10^{-11} \text{m}}{L} \right)^2 \text{Hz}^{-1},
\end{equation}
where $L = 2.5 \cdot 10^9 \ \text{m}$%
\footnote{In~\cite[p. 12]{gowling_lisa_2021} there is a typo in the value of $L$. This thesis uses the correct value from~\cite{smith_lisa_2019}.}
is the LISA constellation arm length.
The single test mass acceleration noise target for LISA is
\cites[eq. 3.3]{gowling_lisa_2021}[eq. 52-53]{smith_lisa_2019}
\begin{equation}
P_\text{acc} = \left( \frac{3 \cdot 10^{-15} \frac{m}{s^2}}{(2\pi f)^2 L} \right)^2 \left( 1 + \left( \frac{0.4 \text{mHz}}{f} \right)^2 \right) \text{Hz}^{-1}.
\end{equation}
The transfer frequency is the inverse of the time it takes for light to be sent between the LISA satellites,
\begin{equation}
f_t \equiv \frac{c}{2 \pi L},
\end{equation}
where $c$ is the speed of light.
The modulation caused by one round trip of a signal along a link is given by
\cite[p. 12]{gowling_lisa_2021}
\begin{equation}
W(f,f_t) = 1 - \exp \left( \frac{-2if}{f_t} \right).
\end{equation}
Using these we can express the instrument noise in the A and E channels of LISA as
\cites[eq. 3.4]{gowling_lisa_2021}[eq. 57]{smith_lisa_2019}
\begin{equation}
N_A = N_E = \left(
\left(4 + 2 \cos \left( \frac{f}{f_t} \right) \right) P_\text{oms} +
8 \left( 1 + \cos \left( \frac{f}{f_t} \right) + \cos^2 \left( \frac{f}{f_t} \right) \right) P_\text{acc}
\right) |W|^2.
\end{equation}
The gravitational wave response function for the A and E channels is known only numerically, but can be approximated as
\cites[eq. 3.6]{gowling_lisa_2021}[eq. 32]{smith_lisa_2019}
\begin{equation}
\mathcal{R}_A^\text{Fit} = \mathcal{R}_E^\text{Fit} \approx \frac{9}{20} |W|^2 \left( 1 + \left( \frac{3f}{4f_t} \right)^2 \right)^{-1}.
\end{equation}

The noise power spectral density (PSD) is defined as
\cite[eq. 3.1]{gowling_lisa_2021}
\begin{equation}
S(f) \equiv \frac{N(f)}{\mathcal{R}(f)}.
\end{equation}
This can be converted to the fractional energy density power spectrum with
\cites[eq. 3.8]{gowling_lisa_2021}[eq. 59]{smith_lisa_2019}
\begin{equation}
\Omega_\text{ins} = \frac{4 \pi^2}{3 H_0^2} f^3 S_A(f).
\end{equation}
Now that we know the power spectra for both the GW signal and noise, we can compute the signal-to-noise ratio with
\cites[eq. 50]{smith_lisa_2019}[eq. 33]{caprini_detecting_2020}[eq. 21]{thrane_sensitivity_2013}
\begin{equation}
\text{SNR} = \sqrt{t_\text{meas} \int_{f_\text{min}}^{f_\text{max}} df \frac{\Omega_\text{signal}(f)^2}{\Omega_\text{noise}(f)^2}},
\end{equation}
where $t_\text{meas}$ is the total measurement time.
Now we have the means to determine whether a phase transition with a particular fluid profile can be detected by LISA.

\chapter{Simulation tools}
\label{ch:simulation}
The equations for the fluid dynamics and gravitational wave production have no analytical solutions in the general case.
Therefore, numerical solutions are required.
To create these solutions, Hindmarsh et al. have developed the phase transition simulation software PTtools~\cite{pttools}.
In this thesis PTtools has been extended to account for speeds of sound that differ from the bag model of section~\ref{bag_model}.
This has enabled PTtools to simulate bubbles using models beyond the bag model,
such as the constant sound speed model of section~\ref{const_cs}.
The previous fluid profile solver relied heavily on analytical shortcuts specific to the bag model,
and the surrounding higher-level functionality was built with this assumption in mind.
Therefore, enabling the support for more complex models required a nearly complete rewrite and significant extension of the entire simulation software.
When also counting the speedup optimisations done by the author as preparations for this update,
these changes resulted in PTtools growing more than an order of magnitude in terms of the lines of code.

\section{Overview of PTtools}
PTtools~\cite{pttools} is a simulation software for modeling the velocity and enthalpy profile of the fluid shell of a single bubble,
and for converting the profile to the velocity spectrum and gravitational wave spectrum using the sound shell model.
PTtools is a
\href{https://www.python.org/}{Python}
library, which uses
\href{https://numpy.org/}{Numpy}
and
\href{https://scipy.org/}{SciPy}
for the numerical simulations,
\href{https://numba.pydata.org/}{Numba}
for speeding up the computations and
\href{https://matplotlib.org/}{Matplotlib}
and
\href{https://plotly.com/}{Plotly}
for plotting.
For PTools and its full documentation, please see~\cite{pttools}.

To use PTtools, the user has to first specify the equation of state.
They can either use the provided bag model (\verb|BagModel|) or constant sound speed model (\verb|ConstCSModel|),
or they can specify their own model by either inheriting from the provided \texttt{Model} class and providing the equation of state analytically.
Another option is to inherit from the provided \verb|ThermoModel| class and provide two of the degrees of freedom $g_p(T,\phi)$ of eq.~\eqref{eq:p_general}, $g_e(T,\phi)$ of~\eqref{eq:e_general} and $g_s(T,\phi)$ of~\eqref{eq:s_general}, and the potential $V(T,\phi)$ of~\eqref{eq:p_general}.
This way the user can let PTtools take care of constructing the equation of state from these parameters.
PTtools then runs various validity checks for the model to ensure that a phase transition will occur and proceed in the system.
For the phase transition to proceed, there has to be a release of energy from the field to the fluid,
and therefore $\Delta V = V_s - V_b \geq 0$.
The model also has to have a critical temperature of eq.~\eqref{eq:critical_temp},
under which it is energetically favourable for the system to transition to the new phase.
As a part of the model initialization, PTtools creates a function for the speed of sound $c_s^2$ of the model.
This function is compiled using Numba to achieve sufficient performance for using this function in the bubble solving.

Once the model has been specified,
the user can create a bubble by creating an instance of the \verb|Bubble| class by providing the model, $v_\text{wall}$ and $\alpha_n$.
PTtools then runs various validity checks for the bubble,
including that there exists a nucleation enthalpy $w_n$ of eq.~\eqref{eq:wn}, a valid solution type of fig.~\ref{fig:solution_types},
and that the nucleation temperature $T_n$ is below the critical temperature.
Unless specifically instructed to delay,
PTtools then numerically finds a solution for the hydrodynamic equations of eq.~\eqref{eq:hydro_param1},~\eqref{eq:hydro_param2}, and~\eqref{eq:hydro_param3} and the bubble wall junction conditions of~\eqref{eq:junction_condition_1} and~\eqref{eq:junction_condition_2},
resulting in a fluid velocity profile for the bubble.
This is also known as solving the bubble.
The \verb|Bubble| object has various methods for extracting key quantities such as the thermodynamic quantities of section~\ref{energy_redistribution}.
PTtools also checks the validity of the solution.

Further details on PTtools are subject to change as the library is being developed.
Please see the PTtools documentation for the latest information.
An example on the Python code required for creating the gravitational wave power spectrum from the phase transition parameters is below.

\lstset{breaklines=true}
\begin{lstlisting}[language=Python]
import matplotlib.pyplot as plt

from pttools.bubble import Bubble
from pttools.models import ConstCSModel
from pttools.omgw0 import Spectrum

# Specify the equation of state
const_cs = ConstCSModel(
    a_s=1.5, a_b=1, css2=1/3, csb2=1/3-0.01, V_s=1
)

# Create a bubble and solve its fluid profile
bubble = Bubble(const_cs, v_wall=0.5, alpha_n=0.2)
bubble.plot()

# Compute gravitational wave velocity and power spectra for the bubble
spectrum = Spectrum(bubble, r_star=0.1)
spectrum.plot_multi()

plt.show()
\end{lstlisting}

\section{Bubble solver}
The bubble solver is the numerical simulation that converts the parameters that describe the phase transition, $v_{\text{wall}}$, $\alpha_n$ and the equation of state, to the fluid velocity and enthalpy profiles $v(\xi)$ and $w(\xi)$.
The bubble solver consists of three steps:
1) preparatory steps that provide initial values for a numerical solver,
2) the numerical solver itself, and
3) post-processing to provide output data in a consistent form.
For the bag model the user can also choose to use the previous version of the solver,
which uses several analytical shortcuts based on the assumption that $c_s^2 = \frac{1}{3}$ and is therefore significantly faster.
This solver can be enabled by calling \verb|Bubble(model, v_wall, alpha_n, use_bag_solver=True)|.

The generic solver starts by checking the type of the solution based on the conditions of table~\ref{table:solution_types}.
If the type of the solution cannot be determined automatically, the solver will halt and request the user to provide the type of the solution.
Then the solver will use the bag model to load reference values for $w_+$ and $w_-$ for the given $v_\text{wall}$ and $\alpha_n$.
These reference values will be used as the starting point for the numerical solver.
If the reference values have not been precomputed,
they will be computed and saved to disk at this point.
If there is no reference data for the given parameters,
as is the case for deflagrations with high $\alpha_n$ and low $v_\text{wall}$ for which there is no bag model solution,
then an arbitrary guess of $w_- = 0.3 w_n$ will be used as the starting point for the solver (subject to change),
and the starting guess of the bubble junction condition solver for $w_+$ will be computed based on the bag model equations.
Finally, the Chapman-Jouguet speed of~\eqref{eq:chapman_jouguet} is computed.

Once these preparations have been done,
the solver chooses an algorithm specific to the type of the solution.
Detonations are the simplest case.
Since the fluid is stationary outside the wall,
the junction conditions can be solved directly using
$\tilde{v}_+ = v_\text{wall}$ and $w_+ = w_n$.
Then the solver integrates from $(v_-, w_-)$ to the fixed point at $(\xi=c_{s,b}, v=0)$.
The fluid inside the fixed point is stationary.
There is also another fixed point at $(\xi=1, v=1)$, but we choose the direction of $\tau$ in eq.~\eqref{eq:hydro_param1},~\eqref{eq:hydro_param2} and~\eqref{eq:hydro_param3} so that we integrate in the correct direction.

Subsonic deflagrations are more complicated,
as we cannot compute $v_{-,\text{sh}}, w_{-,\text{sh}}$ from $v_{+,\text{sh}}=0, w_{+,\text{sh}}=w_n$ using the junction conditions,
since we don't know the shock speed beforehand.
For some deflagrations, the shock can also be too small for the numerical accuracy of the junction solver,
since when $v_\text{sh}$ is close to $c_{s,s}$, the fluid profile can vary significantly with small changes of $v_\text{sh}$.
Therefore, we have to start by guessing a $w_-$.
Since the fluid inside the wall is stationary, we know that $\tilde{v}_- = v_{\text{wall}}$.
For a given $(v_-, w_-)$ we can solve the junction conditions, giving $(v_+, w_+)$.
From these we can integrate until we encounter the shock.
However, we cannot directly compute $v_\text{sh}(\xi_\text{sh})$, as it's dependent on $w_{-,\text{sh}}$.
Therefore we have to evaluate $v_\text{sh}(\xi_\text{sh}, w_{-,\text{sh}})$ for each $(\xi, w)$ on the integrated curve to see
when the curve encounters the shock.
This is accomplished by a binary search.
Once we know $\xi_\text{sh}$, we can solve the junction conditions with $v_{+,\text{sh}} = 0$, giving us a $w_{+,\text{sh}}$.
Then we compare this to the given $w_n$.
If they don't match, we adjust our guess for $w_-$ and start again until we have $w_{+,\text{sh}} = w_n$.

Hybrids are the most complicated case, as neither $v_+$ nor $v_-$ is known beforehand.
Therefore we start by guessing a $w_-$, and we know that $\tilde{v}_- = c_s(w_-, \phi_-)$.
Then we can solve the junction conditions to get $(v_+, w_+)$.
With these we can integrate until we encounter the shock.
Then we perform the rest of the steps as for deflagrations, and iterate until we have $w_{+,\text{sh}} = w_n$.
Once this is found, we integrate from our starting point $(v_-, w_-)$ to the fixed point to get the detonation-like tail,
resulting in the full hybrid solution.

For all solution types, once we have the solution,
we add points corresponding to $(\xi=0, w_\text{center})$ and $(\xi=1, w_n)$
so that the solution covers the full range $\xi \in [0, 1]$.
It should be noted, that the resulting solution does not have a fixed step for $\xi$,
and therefore one has to be careful when computing integrals or Fourier transforms of the solution.

The \verb|Bubble| class performs various checks on the solution.
It checks that
1) $\alpha_+(w_+, w_-, \tilde{v}_+)$ can be computed,
2) entropy fluxes across the wall and their difference are non-negative,
3) the total change in the volume-averaged entropy density is non-negative, and
4) that energy is conserved by $\kappa + \omega \approx 1$~\eqref{eq:kappa_omega}.
If any of these checks fail, the solution is marked to have an error.

Once the bubble is solved, the \verb|Bubble| object can be queried for various quantities,
including but not limited to the fluid velocities in both the plasma and wall frames,
the corresponding velocities for the shock,
enthalpies at the wall and at the shock,
various thermodynamic quantities in both bubble volume averaged and volume-averaged forms,
such as the entropy density and entropy fluxes,
kinetic energy density~\eqref{eq:kinetic_energy_density},
kinetic energy fraction~\eqref{eq:kinetic_energy_fraction},
thermal energy density~\eqref{eq:thermal_energy_density},
trace anomaly $\theta$~\eqref{eq:theta},
mean energy density $\bar{e} = e_n$~\eqref{eq:e_conservation},
mean enthalpy density $\bar{w}$~\eqref{eq:wbar},
$\kappa$~\eqref{eq:kappa_omega},
$\kappa_{\bar{\theta}_+}$~\eqref{eq:kappa_thetabar_plus},
$\kappa_{\bar{\theta}_n}$~\eqref{eq:kappa_thetabar_n},
$\omega$~\eqref{eq:kappa_omega},
mean adiabatic index $\Gamma$~\eqref{eq:mean_adiabatic_index} and
$\bar{U}_f^2$~\eqref{eq:ubarf2}.

\section{Spectrum computation}
Once the fluid velocity profile of a bubble is solved as above,
it can be converted to the velocity spectrum and gravitational wave spectrum by constructing a \verb|Spectrum| object.
The spectrum computation is based on the Sound Shell Model~\cite{hindmarsh_gw_pt_2019}.
First, the spectral density of the plane wave components of the velocity field $P_v(q)$ is computed using eq.~\eqref{eq:spec_den_v}.
This is converted to the velocity power spectrum $\mathcal{P}_{\tilde{v}}$ using eq.~\eqref{eq:pow_v}.
The spectral density of the plane wave components of the velocity field $P_v(q)$ is also used to compute the spectral density of gravitational waves $\tilde{P}_{\text{gw}}(y)$ of eq.~\eqref{eq:spectral_density}.
This is converted to the gravitational wave power spectrum $\mathcal{P}_{\text{gw}}$ using~\eqref{eq:gw_pow_spec3}.

\section{Parallel computing}
PTtools provides an interface for creating multiple bubbles and computing quantities from them in parallel on multiple CPU cores,
despite being Python-based software.
This is made possible by the
\href{https://docs.python.org/3/library/multiprocessing.html}{\texttt{multiprocessing}}
module of the Python standard library.
The spectrum computation of a single bubble can also take advantage of multiple CPU cores thanks to Numba parallelism.
PTtools has various utilities built on top of this that simplify the parallel generation of bubbles.
An example of a parallel program is provided below.

\begin{lstlisting}[language=Python]
import numpy as np

from pttools.analysis import BubbleGridVWAlpha
from pttools.bubble import Bubble
from pttools.models import BagModel


def compute(bubble: Bubble):
	if bubble.no_solution_found or bubble.solver_failed:
		return np.nan, np.nan
	return bubble.kappa, bubble.omega

compute.return_type = (float, float)


v_walls = np.linspace(0.1, 0.9, 5)
alpha_ns = np.linspace(0.1, 0.3, 5)
model = BagModel(a_s=1.1, a_b=1, V_s=1)
grid = BubbleGridVWAlpha(model, v_walls, alpha_ns, compute)
bubbles = grid.bubbles
kappas = grid.data[0]
omegas = grid.data[1]
\end{lstlisting}

\chapter{Results}
\label{ch:results}
\begin{figure}[ht!]
\centering
\includegraphics[width=\textwidth]{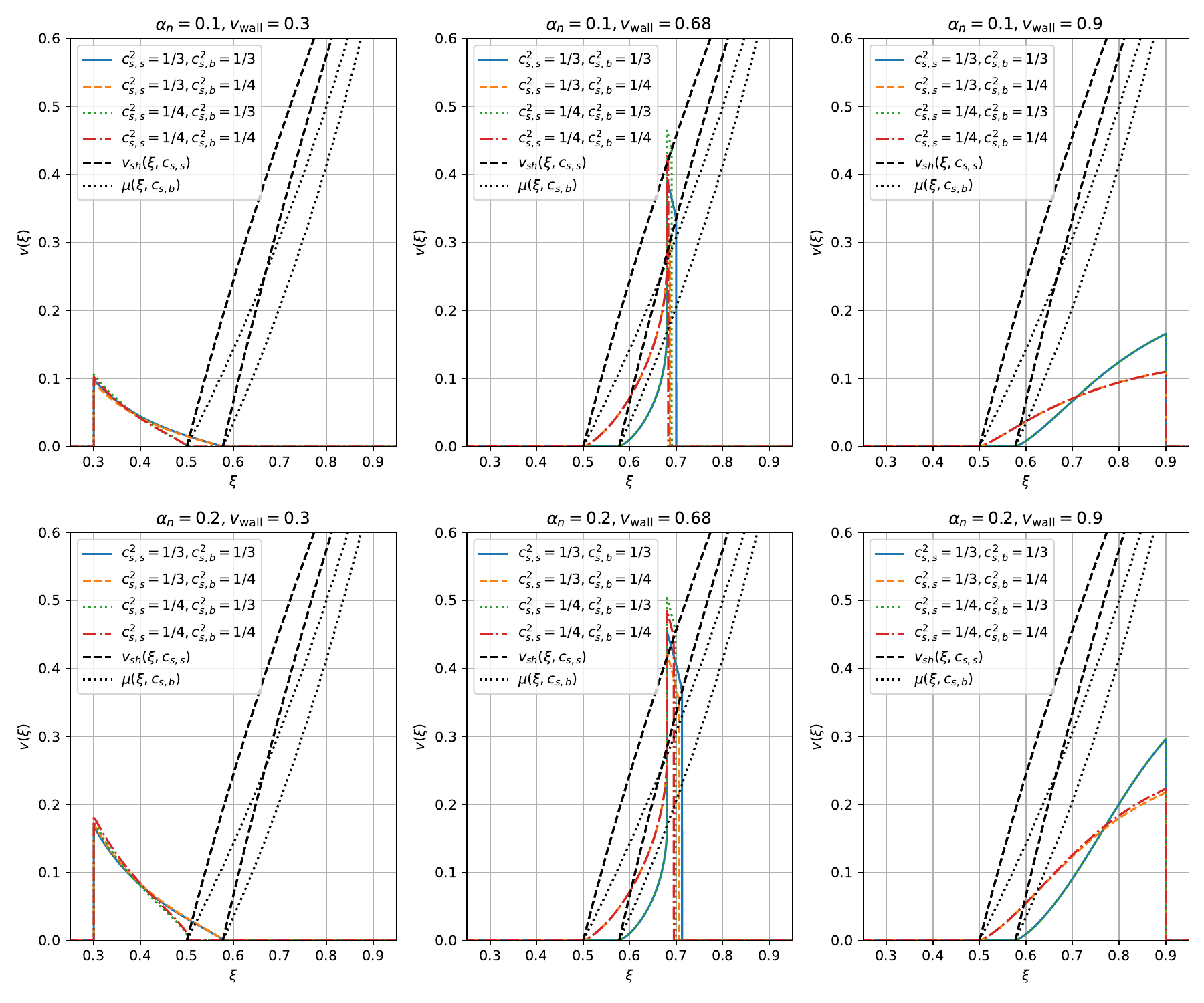}
\caption{Self-similar fluid profiles with four sound speed combinations, three different wall speeds $v_\text{wall}$ and two transition strengths $\alpha_n$}
\label{fig:fluid_profiles}
\end{figure}

\begin{figure}[ht!]
\centering
\includegraphics[width=\textwidth]{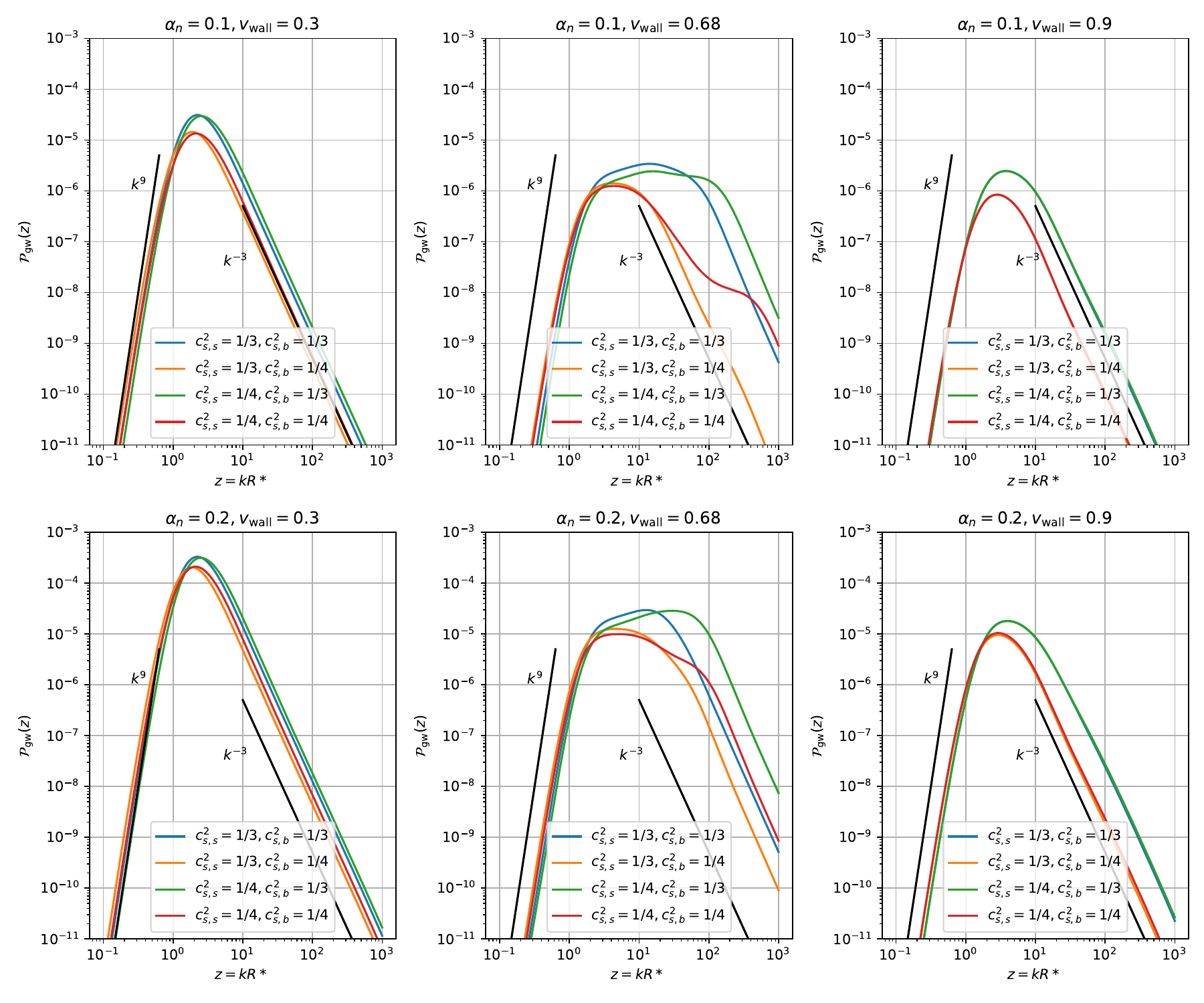}
\caption{Gravitational wave power spectra computed with the sound shell model from the fluid profiles of fig.~\ref{fig:fluid_profiles}}
\label{fig:gw_spectra}
\end{figure}

\begin{figure}[ht!]
\centering
\includegraphics[width=\textwidth]{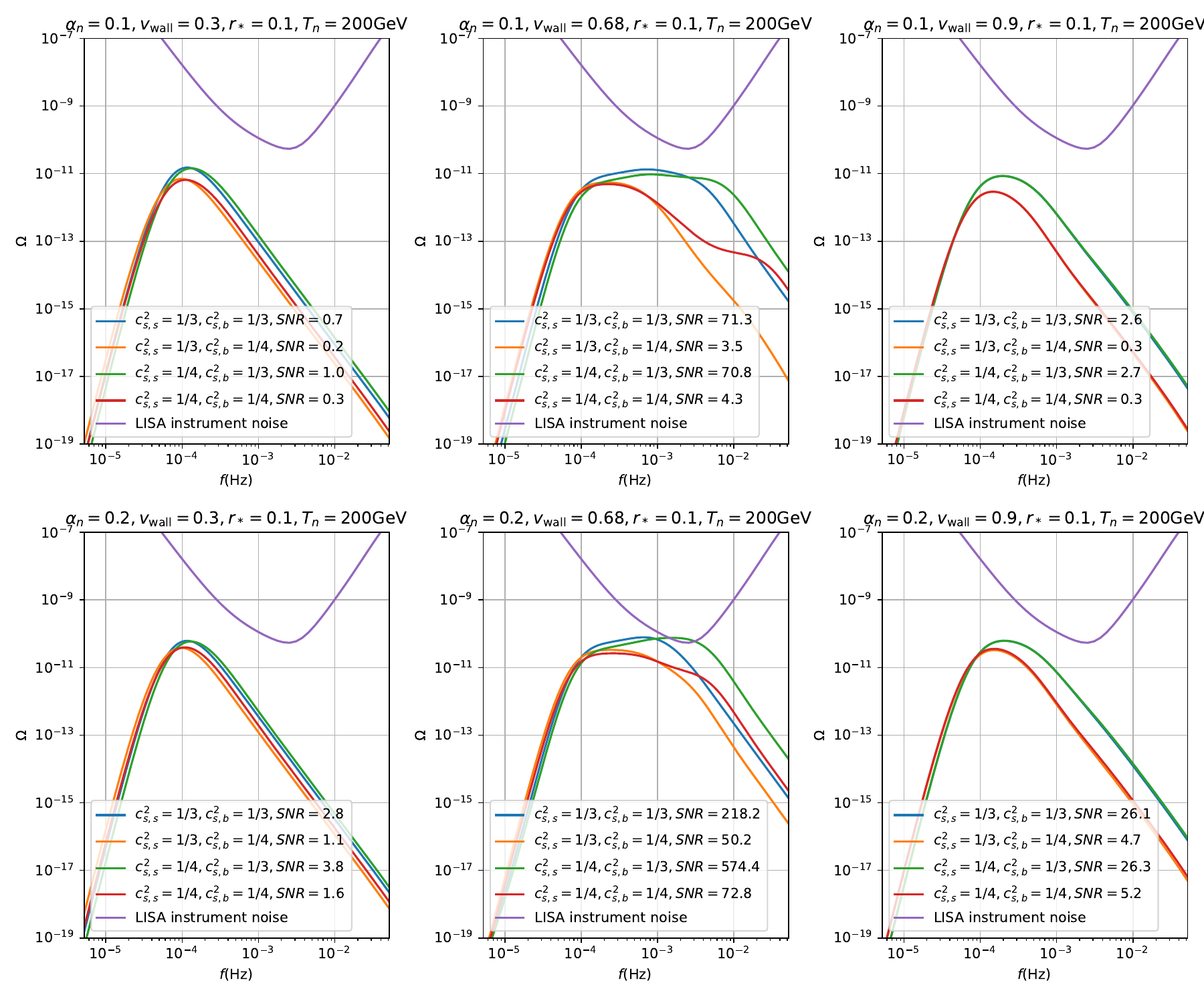}
\caption{Scaled gravitational wave power spectra today, $\Omega_{gw,0}$, of the gravitational wave power spectra of fig.~\ref{fig:gw_spectra}}
\label{fig:omgw0}
\end{figure}

The constant $c_s$ model was chosen as the model to test PTtools with,
as the model is the next logical extension of the bag model,
and there is some reference data available from~\cites{giese_2020}{giese_2021}.
Figure~\ref{fig:fluid_profiles} demonstrates solutions for three different wall speeds $v_\text{wall}$
and two different transition strengths $\alpha_n$ of eq.~\eqref{eq:alpha_n}.
Instead of using only the bag model which has the squared sound speeds $c_{s,s}^2 = c_{s,b}^2 = \frac{1}{3}$,
each plot has four curves, each corresponding to a different combination of the squared sound speeds
$c_{s,s}^2 \in \{ \frac{1}{3}, \frac{1}{4} \}, c_{s,b}^2 \in \{ \frac{1}{3}, \frac{1}{4} \}$.
The corresponding gravitational wave spectra are plotted in figure~\ref{fig:gw_spectra}.
Converting these to the frequencies and amplitudes today results in figure~\ref{fig:omgw0},
which also contains the LISA instrument noise spectrum as a reference.

These figures demonstrate that the sound speed can have a profound effect on the resulting gravitational wave spectrum.
To understand how the sound speed affects the gravitational wave spectrum,
we have to first study the fluid shells of figure~\ref{fig:fluid_profiles}.
There are two curves that constrain the shape of the fluid shells.
The solid black curves are the shock velocities $v_\text{sh}(\xi,c_{s,s})$ that are determined by $c_{s,s}$.
In general the shock velocity is also dependent on the enthalpy density $w$, since $c_s = c_s(w,\phi)$,
but in the constant sound speed model the sound speed is independent of the enthalpy,
as it is a constant for each phase.
Therefore, we can plot the shock velocities as curves in the 2D plots
instead of having to plot them as surfaces that are functions of $(\xi,w)$ in a 3D plot.
The dotted black curves are the $\mu(\xi,c_{s,b})$ curves of eq.~\eqref{eq:mu} that constrain the maximum fluid velocity inside the bubble.
They are dependent on the choice of $c_{s,b}$.
The $v_\text{sh}$ curves encounter $v=0$ at $\xi = c_{s,s}$ and the $\mu$ curves at $\xi = c_{s,b}$.

For all solution types, adjusting either of the sound speeds affects the degrees of freedom
and therefore the pressure $p$ and enthalpy density $w$ for that phase.
This affects the solution of the bubble wall junction conditions of eq.~\eqref{eq:junction_condition_1},~\eqref{eq:junction_condition_2} and therefore $v(\xi_\text{wall})$,
which is also the peak of the fluid velocity profile.
This effect can be seen at the top-left figure, where all four fluid shells have different $v(\xi_\text{wall})$.
The change caused by adjusting the sound speed is the most profound for deflagrations when $c_{s,s}$ is decreased below $v_\text{wall}$, as it causes them to become hybrids, and vice versa.
Similarly, changing $c_{s,b}$ so that the Chapman-Jouguet speed of~\eqref{eq:chapman_jouguet} is decreased below $v_\text{wall}$ converts a hybrid to a detonation, and vice versa.

For deflagrations and hybrids, adjusting $c_{s,s}$ affects the location of the shock and therefore the thickness of the fluid shell.
It also affects the ODE group of eq.~\eqref{eq:hydro_param1},~\eqref{eq:hydro_param2},~\eqref{eq:hydro_param3}
and therefore the shape of the fluid shell in front of the bubble wall.
These effects can be seen at the top-left and bottom-left figures, where the curves with the same $c_{s,s}$ are grouped together.
Correspondingly for hybrids, adjusting $c_{s,b}$ affects $\mu(\xi_\text{wall},c_{s,b})$,
which is the point from which the integration of the detonation-like part of the fluid shell starts.
This can be seen in the two figures at the middle, where the detonation-like tails of the curves with the same $c_{s,b}$ are grouped together.
And for both detonations and hybrids, adjusting $c_{s,b}$ affects the shape of the fluid shell behind the wall.

These differences in the shapes of the fluid shells carry over to the gravitational wave spectra of fig.~\ref{fig:gw_spectra}
and eventually of the gravitational wave spectra today in fig.~\ref{fig:omgw0}.
The sine transformation of eq.~\eqref{eq:ssm_f} and~\eqref{eq:ssm_l} in the process converts the fluid shells of fig.~\ref{fig:fluid_profiles}
to the gravitational wave spectra of~\ref{fig:gw_spectra} has the same basic properties as a Fourier transform.
Therefore, by comparing the fluid velocity profiles and the gravitational wave spectra we can identify a few general correlations.
The thinner the fluid shell is, the broader the gravitational wave spectrum.
And the higher the fluid velocities are in the fluid shell, the higher is the intensity of the gravitational waves.
Thin shells also have more of the higher frequencies.
These effects can be seen in the case of thin hybrids of the middle figures.
Hybrids consist of two sections with significantly differing characteristics.
Therefore, the hybrids and especially the thin hybrids with high fluid velocities in front of the wall have a gravitational wave spectrum with two distinct contributions.
This results in a significantly higher gravitational wave spectrum in the higher frequencies,
resulting in a significantly higher signal-to-noise ratio (SNR),
which distinguishes them from the detonations and deflagrations.
Outside the region of the peak, the gravitational wave spectra follow a $k^9$ power law at low $z$,
and $k^{-3}$ at high $z$.

The signal-to-noise ratios of the example spectra in figure~\ref{fig:omgw0}
are entirely below the LISA noise curve except for the one curve with
$c_{s,s} = \frac{1}{4}, c_{s,b} = \frac{1}{3}, \alpha_n = 0.2, v_\text{wall} = 0.68$,
but the signal-to-noise ratios above unity by an order of magnitude or two for many of the curves.
This is thanks to the integrated nature of the signal,
which makes the overall signal distinguishable from the noise power spectrum
even though the power at individual frequencies is below the noise.
\cite{thrane_sensitivity_2013}

\begin{figure}[ht!]
\centering
\includegraphics[width=\textwidth]{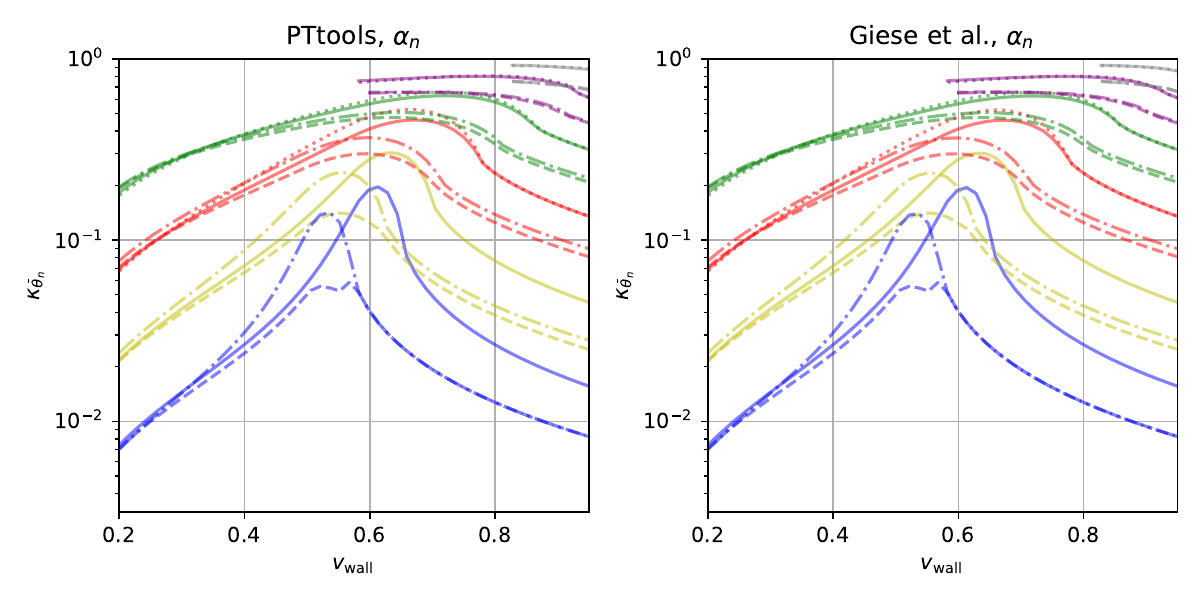}
\caption{Comparison of $\kappa$ values by~\cite[fig. 2]{giese_2021} and PTtools}
\label{fig:kappa_giese}
\end{figure}
\begin{figure}[ht!]
\centering
\includegraphics[width=\textwidth]{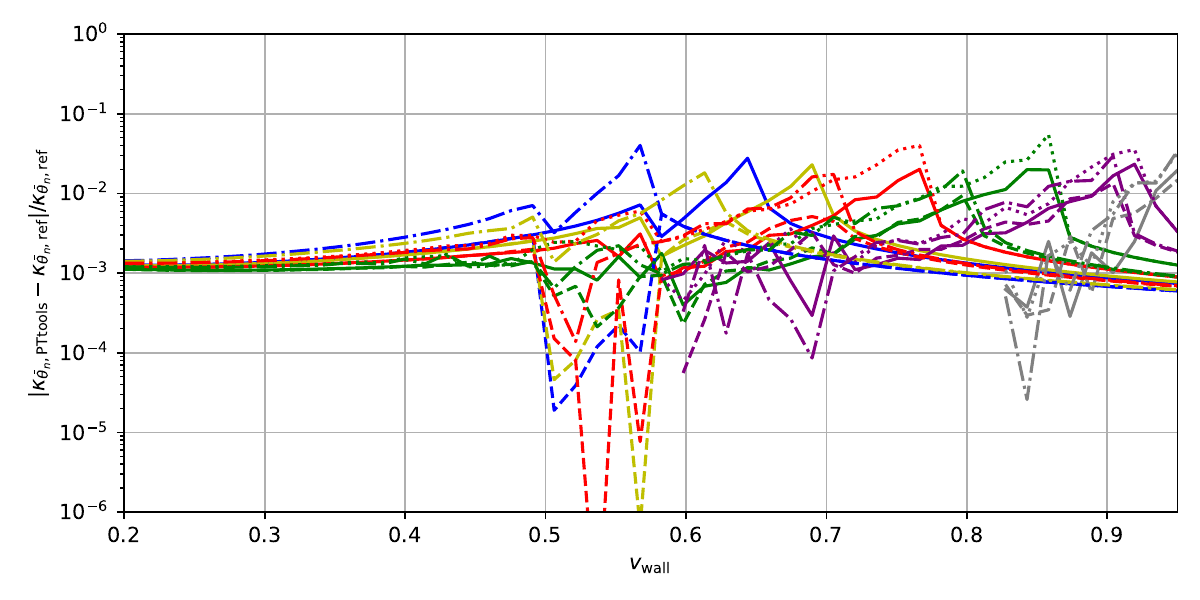}
\caption{Relative difference of the $\kappa$ values by~\cite[fig. 2]{giese_2021} and PTtools}
\label{fig:kappa_giese_diff}
\end{figure}

To test the precision and reliability of the results provided by PTtools,
we compared to the results given by the code in  Giese et al., 2021~\cite[fig. 2]{giese_2021}.
The reference fluid profiles are generated by the code in the article,
and the $\kappa$ of eq.~\eqref{eq:kappa_omega} is computed by PTtools.
The resulting $\kappa$ values are in fig.~\ref{fig:kappa_giese},
and the relative differences in fig.~\ref{fig:kappa_giese_diff}.
The colors from blue to gray correspond to $\alpha = 0.01, \ 0.03, \ 0.1, \ 0.3, \ 1, \ 3$.
For each color, the solid line has $c_{s,s}^2 = \frac{1}{3}, \ c_{s,b}^2 = \frac{1}{3}$,
the dashed line has $c_{s,s}^2 = \frac{1}{3}, \ c_{s,b}^2 = \frac{1}{4}$,
the dotted line has $c_{s,s}^2 = \frac{1}{4}, \ c_{s,b}^2 = \frac{1}{3}$
and the dot-dashed line has $c_{s,s}^2 = \frac{1}{4}, \ c_{s,b}^2 = \frac{1}{4}$.

As we can see by the visually identical results in fig.~\ref{fig:kappa_giese}
and the small scale of the relative differences in fig.~\ref{fig:kappa_giese_diff},
the fluid profile simulations give nearly identical results.
This demonstrates that PTtools is a reliable fluid profile simulator for a variety of
sound speeds $c_s$, phase transition strengths $\alpha_n$ and wall speeds $v_\text{wall}$.
The differences are due to the differences in the number of the integration points chosen for the curve
and the different junction condition solver and ODE integration algorithms.

For $c_{s,s}=\frac{1}{4}, \ c_{s,b}=\frac{1}{3}$ and $\alpha_n = 0.01, \ 0.03$, there is no curve,
as the $\alpha_n$ is below the theoretical minimum of $\alpha_n \approx 0.067$ for the model.
This is recognised by the Giese et al. solver as well,
which is why it lacks the same curves.

The code by Giese et al. is specific to the constant sound speed model,
whereas the PTtools solver is desiged to be generic in such a way that
it can solve bubbles with arbitrary $c_s(T,\phi)$.
The PTtools code also enables easy extraction of various thermodynamic quantities
and the gravitational wave spectra.
Therefore, now that we know that PTtools works reliably for the constant sound speed model,
the next step is to test it with more complex models.

\chapter{Conclusion}
\label{ch:conclusion}
In this thesis PTtools has been expanded from a compact simulation script based on the bag model
to a comprehensive framework for simulating the fluid velocity profiles of first-order phase transitions in the early universe
with temperature- and phase-dependent speeds of sound $c_s(T,\phi)$,
and their gravitational wave spectra based on the Sound Shell Model,
including the conversion to the gravitational spectra today.

PTtools was tested with the constant sound speed model due to the availability of reference data by Giese et al.~\cite{giese_2021},
and it was shown to work reliably for a broad range of the combinations of the sound speeds $c_{s,s}$ and $c_{s,b}$, the phase transition strength $\alpha_n$ and the wall speed $v_\text{wall}$.
The PTtools fluid velocity profile solver has been updated beyond that of the available references in such a way
that it supports arbitrary particle physics models with a temperature-dependent sound speed $c_s(T,\phi)$,
as long as they provide $V_s(T,\phi), V_b(T,\phi)$ and two of $g_p(T,\phi), g_e(T,\phi), g_s(T,\phi)$ or two of $p(T,\phi), e(T,\phi), s(T,\phi)$, but preferably all three for numerical precision.
This enables the easy integration of models developed by other researchers,
and therefore the comparison of the gravitational wave spectra of these models.
This is also to the author's best knowledge the first time
that phase transitions in the early universe have been simulated with temperature-dependent speeds of sound
without resorting to time-consuming 3D simulations.

The numerical results of the fluid shell solver have been demonstrated to be consistent with a reference
for a broad range of parameters.
However, some special cases are still challenging.
The most notable of these are very thin hybrid shells near the Chapman-Jouguet speed $v_\text{wall} = v_{CJ}$,
where the thickness of the fluid shell in front of the wall is comparable to the minimum step possible for the ODE solvers.
This limits the precision of the overall solver in these special cases.
However, these special cases compose a very small section of the overall parameter space,
and for the vast majority of the parameter space,
the PTtools solver produces reliable results.

The performance of PTtools is also notable.
The creation of the gravitational wave spectrum today from the phase transition parameters takes less than a second per bubble on a laptop,
and generating all the figures in this thesis takes less than five minutes on a workstation.
This is in a stark contrast with the 3D hydrodynamic simulations that require the computational power of supercomputers.
This demonstrates that the Sound Shell Model is an essential tool in investigating
the effects of the phase transition parameters on the resulting gravitational wave spectrum.

This master's thesis lays the groundwork for the author's PhD thesis on the topic.
Now there is a framework for determining the gravitational wave spectrum from the phase transition parameters,
including temperature-dependent speed of sound.
Determining the parameters of a phase transition from LISA data will require solving the inverse problem:
what are the phase transition parameters based on the gravitational wave spectrum?
This will require novel tools such as machine learning or Markov chain Monte Carlo simulations.
There is also potential for integrating PTtools with the existing web-based simulation utility
PTPlot~\cites{ptplot}{hindmarsh_shape_2017}{caprini_detecting_2020}
to create a comprehensive but easy-to-use web-based utility for researchers to simulate phase transitions with various parameters.
Further possibilities for extending PTtools itself include integrating
the low frequency spectral density function by Giombi et al.~\cite[eq. 3.6]{giombi_cs_2024}.
These topics will be investigated in the author's PhD studies.

PTtools is available as open source on GitHub at~\cite{pttools}.
This enables the research community to integrate their various particle physics models,
and to simulate the resulting gravitational wave spectra.
This will provide the LISA simulation pipeline with various waveforms of interest,
and if one of them is eventually found in LISA data,
this will result in a groundbreaking discovery
that will point the direction for the development of particle physics beyond the Standard Model.

\cleardoublepage 
\phantomsection

\addcontentsline{toc}{chapter}{\bibname} 

\printbibliography

@book{landau_fluid_1987,
	location = {Oxford, England ; New York},
	edition = {2nd ed., 2nd English ed., rev},
	title = {Fluid mechanics},
	isbn = {978-0-08-033933-7 978-0-08-033932-0},
	series = {Course of theoretical physics},
	pagetotal = {539},
	number = {v. 6},
	publisher = {Pergamon Press},
	author = {Landau, L. D. and Lifshitz, E. M.},
	date = {1987},
}

@book{schroeder_thermal_2000,
	location = {San Francisco, {CA}},
	title = {An introduction to thermal physics},
	isbn = {978-0-201-38027-9},
	pagetotal = {422},
	publisher = {Addison Wesley},
	author = {Schroeder, Daniel V.},
	date = {2000}
}

@book{maggiore_gw_2008,
	location = {Oxford},
	title = {Gravitational waves},
	isbn = {978-0-19-857074-5 978-0-19-857089-9},
	pagetotal = {1},
	publisher = {Oxford University Press},
	author = {Maggiore, Michele},
	date = {2008},
	note = {{OCLC}: ocn180464569},
}

@book{rezzolla_relativistic_2013,
	location = {Oxford},
	edition = {1. ed},
	title = {Relativistic hydrodynamics},
	isbn = {978-0-19-150991-9 978-0-19-852890-6 978-0-19-880759-9},
	pagetotal = {735},
	publisher = {Oxford Univ. Press},
	author = {Rezzolla, Luciano and Zanotti, Olindo},
	date = {2013},
}

@article{kurki-suonio_supersonic_1995,
	title = {Supersonic deflagrations in cosmological phase transitions},
	volume = {51},
	copyright = {http://link.aps.org/licenses/aps-default-license},
	issn = {0556-2821},
	url = {https://link.aps.org/doi/10.1103/PhysRevD.51.5431},
	doi = {10.1103/PhysRevD.51.5431},
	language = {en},
	number = {10},
	journal = {Physical Review D},
	author = {Kurki-Suonio, H. and Laine, M.},
	month = may,
	year = {1995},
	pages = {5431--5437}
}

@article{moore_pt_1995,
	title = {How fast can the wall move? A study of the electroweak phase transition dynamics},
	volume = {52},
	issn = {0556-2821},
	url = {https://link.aps.org/doi/10.1103/PhysRevD.52.7182},
	doi = {10.1103/PhysRevD.52.7182},
	shorttitle = {How fast can the wall move?},
	pages = {7182--7204},
	number = {12},
	journaltitle = {Physical Review D},
	shortjournal = {Phys. Rev. D},
	author = {Moore, Guy D. and Prokopec, Tomislav},
	date = {1995-12-15},
	langid = {english}
}

@article{kajantie_is_1996,
	title = {Is {There} a {Hot} {Electroweak} {Phase} {Transition} at $m_H \gtrsim m_W$?},
	volume = {77},
	issn = {0031-9007, 1079-7114},
	url = {https://link.aps.org/doi/10.1103/PhysRevLett.77.2887},
	doi = {10.1103/PhysRevLett.77.2887},
	language = {en},
	number = {14},
	journal = {Physical Review Letters},
	author = {Kajantie, K. and Laine, M. and Rummukainen, K. and Shaposhnikov, M.},
	month = sep,
	year = {1996},
	pages = {2887--2890}
}

@article{aoki_order_2006,
	title = {The order of the quantum chromodynamics transition predicted by the standard model of particle physics},
	volume = {443},
	rights = {http://www.springer.com/tdm},
	issn = {0028-0836, 1476-4687},
	url = {https://www.nature.com/articles/nature05120},
	doi = {10.1038/nature05120},
	pages = {675--678},
	number = {7112},
	journaltitle = {Nature},
	shortjournal = {Nature},
	author = {Aoki, Y. and Endrődi, G. and Fodor, Z. and Katz, S. D. and Szabó, K. K.},
	date = {2006-10},
	langid = {english}
}

@article{bodeker_can_2009,
	title = {Can electroweak bubble walls run away?},
	volume = {2009},
	issn = {1475-7516},
	url = {https://iopscience.iop.org/article/10.1088/1475-7516/2009/05/009},
	doi = {10.1088/1475-7516/2009/05/009},
	number = {05},
	journal = {Journal of Cosmology and Astroparticle Physics},
	author = {Bödeker, Dietrich and Moore, Guy D},
	month = may,
	year = {2009},
	pages = {009--009},
}

@article{espinosa_energy_2010,
	title = {Energy budget of cosmological first-order phase transitions},
	volume = {2010},
	issn = {1475-7516},
	url = {https://iopscience.iop.org/article/10.1088/1475-7516/2010/06/028},
	doi = {10.1088/1475-7516/2010/06/028},
	pages = {028--028},
	number = {6},
	journaltitle = {Journal of Cosmology and Astroparticle Physics},
	shortjournal = {J. Cosmol. Astropart. Phys.},
	author = {Espinosa, José R and Konstandin, Thomas and No, José M and Servant, Géraldine},
	date = {2010-06-28}
}

@article{thrane_sensitivity_2013,
	title = {Sensitivity curves for searches for gravitational-wave backgrounds},
	volume = {88},
	copyright = {http://link.aps.org/licenses/aps-default-license},
	issn = {1550-7998, 1550-2368},
	url = {https://link.aps.org/doi/10.1103/PhysRevD.88.124032},
	doi = {10.1103/PhysRevD.88.124032},
	language = {en},
	number = {12},
	journal = {Physical Review D},
	author = {Thrane, Eric and Romano, Joseph D.},
	month = dec,
	year = {2013},
	pages = {124032}
}

@article{leitao_hydrodynamics_2015,
	title = {Hydrodynamics of phase transition fronts and the speed of sound in the plasma},
	volume = {891},
	issn = {05503213},
	url = {https://linkinghub.elsevier.com/retrieve/pii/S0550321314003770},
	doi = {10.1016/j.nuclphysb.2014.12.008},
	language = {en},
	journal = {Nuclear Physics B},
	author = {Leitao, Leonardo and Mégevand, Ariel},
	month = feb,
	year = {2015},
	pages = {159--199}
}

@article{borsanyi_lattice_2016,
	title = {Lattice {QCD} for Cosmology},
	url = {http://arxiv.org/abs/1606.07494},
	author = {Borsanyi, Sz and Fodor, Z. and Kampert, K. H. and Katz, S. D. and Kawanai, T. and Kovacs, T. G. and Mages, S. W. and Pasztor, A. and Pittler, F. and Redondo, J. and Ringwald, A. and Szabo, K. K.},
	date = {2016-06-27},
	eprinttype = {arxiv},
	eprint = {1606.07494}
}

@misc{lisa_2017,
	title = {Laser {Interferometer} {Space} {Antenna}},
	copyright = {arXiv.org perpetual, non-exclusive license},
	url = {https://arxiv.org/abs/1702.00786},
	doi = {10.48550/ARXIV.1702.00786},
	publisher = {arXiv},
	author = {Amaro-Seoane, Pau and Audley, Heather and Babak, Stanislav and Baker, John and Barausse, Enrico and Bender, Peter and Berti, Emanuele and Binetruy, Pierre and Born, Michael and Bortoluzzi, Daniele and Camp, Jordan and Caprini, Chiara and Cardoso, Vitor and Colpi, Monica and Conklin, John and Cornish, Neil and Cutler, Curt and Danzmann, Karsten and Dolesi, Rita and Ferraioli, Luigi and Ferroni, Valerio and Fitzsimons, Ewan and Gair, Jonathan and Bote, Lluis Gesa and Giardini, Domenico and Gibert, Ferran and Grimani, Catia and Halloin, Hubert and Heinzel, Gerhard and Hertog, Thomas and Hewitson, Martin and Holley-Bockelmann, Kelly and Hollington, Daniel and Hueller, Mauro and Inchauspe, Henri and Jetzer, Philippe and Karnesis, Nikos and Killow, Christian and Klein, Antoine and Klipstein, Bill and Korsakova, Natalia and Larson, Shane L and Livas, Jeffrey and Lloro, Ivan and Man, Nary and Mance, Davor and Martino, Joseph and Mateos, Ignacio and McKenzie, Kirk and McWilliams, Sean T and Miller, Cole and Mueller, Guido and Nardini, Germano and Nelemans, Gijs and Nofrarias, Miquel and Petiteau, Antoine and Pivato, Paolo and Plagnol, Eric and Porter, Ed and Reiche, Jens and Robertson, David and Robertson, Norna and Rossi, Elena and Russano, Giuliana and Schutz, Bernard and Sesana, Alberto and Shoemaker, David and Slutsky, Jacob and Sopuerta, Carlos F. and Sumner, Tim and Tamanini, Nicola and Thorpe, Ira and Troebs, Michael and Vallisneri, Michele and Vecchio, Alberto and Vetrugno, Daniele and Vitale, Stefano and Volonteri, Marta and Wanner, Gudrun and Ward, Harry and Wass, Peter and Weber, William and Ziemer, John and Zweifel, Peter},
	year = {2017},
	note = {Version Number: 3}
}

@article{bodeker_electroweak_2017,
	title = {Electroweak bubble wall speed limit},
	volume = {2017},
	copyright = {http://iopscience.iop.org/info/page/text-and-data-mining},
	issn = {1475-7516},
	url = {https://iopscience.iop.org/article/10.1088/1475-7516/2017/05/025},
	doi = {10.1088/1475-7516/2017/05/025},
	number = {05},
	journal = {Journal of Cosmology and Astroparticle Physics},
	author = {Bödeker, Dietrich and Moore, Guy D.},
	month = may,
	year = {2017},
	pages = {025--025},
}

@article{hindmarsh_shape_2017,
	title = {Shape of the acoustic gravitational wave power spectrum from a first order phase transition},
	volume = {96},
	copyright = {https://link.aps.org/licenses/aps-default-license},
	issn = {2470-0010, 2470-0029},
	url = {https://link.aps.org/doi/10.1103/PhysRevD.96.103520},
	doi = {10.1103/PhysRevD.96.103520},
	language = {en},
	number = {10},
	journal = {Physical Review D},
	author = {Hindmarsh, Mark and Huber, Stephan J. and Rummukainen, Kari and Weir, David J.},
	month = nov,
	year = {2017},
	pages = {103520}
}

@article{smith_lisa_2019,
	title = {{LISA} for cosmologists: {Calculating} the signal-to-noise ratio for stochastic and deterministic sources},
	volume = {100},
	issn = {2470-0010, 2470-0029},
	shorttitle = {{LISA} for cosmologists},
	url = {https://link.aps.org/doi/10.1103/PhysRevD.100.104055},
	doi = {10.1103/PhysRevD.100.104055},
	language = {en},
	number = {10},
	journal = {Physical Review D},
	author = {Smith, Tristan L. and Caldwell, Robert R.},
	month = nov,
	year = {2019},
	pages = {104055}
}

@article{mazumdar_review_2019,
	title = {Review of cosmic phase transitions: their significance and experimental signatures},
	volume = {82},
	issn = {0034-4885, 1361-6633},
	url = {https://iopscience.iop.org/article/10.1088/1361-6633/ab1f55},
	doi = {10.1088/1361-6633/ab1f55},
	shorttitle = {Review of cosmic phase transitions},
	pages = {076901},
	number = {7},
	journaltitle = {Reports on Progress in Physics},
	shortjournal = {Rep. Prog. Phys.},
	author = {Mazumdar, Anupam and White, Graham},
	date = {2019-07-01}
}

@article{hindmarsh_gw_pt_2019,
	title = {Gravitational waves from first order cosmological phase transitions in the Sound Shell Model},
	volume = {2019},
	issn = {1475-7516},
	url = {https://iopscience.iop.org/article/10.1088/1475-7516/2019/12/062},
	doi = {10.1088/1475-7516/2019/12/062},
	pages = {062--062},
	number = {12},
	journaltitle = {Journal of Cosmology and Astroparticle Physics},
	shortjournal = {J. Cosmol. Astropart. Phys.},
	author = {Hindmarsh, Mark and Hijazi, Mulham},
	date = {2019-12-23}
}

@article{caprini_detecting_2020,
	title = {Detecting gravitational waves from cosmological phase transitions with {LISA}: an update},
	volume = {2020},
	issn = {1475-7516},
	url = {https://iopscience.iop.org/article/10.1088/1475-7516/2020/03/024},
	doi = {10.1088/1475-7516/2020/03/024},
	shorttitle = {Detecting gravitational waves from cosmological phase transitions with {LISA}},
	pages = {024--024},
	number = {3},
	journaltitle = {Journal of Cosmology and Astroparticle Physics},
	shortjournal = {J. Cosmol. Astropart. Phys.},
	author = {Caprini, Chiara and Chala, Mikael and Dorsch, Glauber C. and Hindmarsh, Mark and Huber, Stephan J. and Konstandin, Thomas and Kozaczuk, Jonathan and Nardini, Germano and No, Jose Miguel and Rummukainen, Kari and Schwaller, Pedro and Servant, Geraldine and Tranberg, Anders and Weir, David J.},
	date = {2020-03-10}
}

@article{giese_2020,
	title = {Model-independent energy budget of cosmological first-order phase transitions -- A sound argument to go beyond the bag model},
	volume = {2020},
	issn = {1475-7516},
	url = {https://iopscience.iop.org/article/10.1088/1475-7516/2020/07/057},
	doi = {10.1088/1475-7516/2020/07/057},
	pages = {057--057},
	number = {7},
	journaltitle = {Journal of Cosmology and Astroparticle Physics},
	shortjournal = {J. Cosmol. Astropart. Phys.},
	author = {Giese, Felix and Konstandin, Thomas and de Vis, Jorinde van},
	date = {2020-07-27},
}

@article{giese_2021,
	title = {Model-independent energy budget for {LISA}},
	volume = {2021},
	issn = {1475-7516},
	url = {https://iopscience.iop.org/article/10.1088/1475-7516/2021/01/072},
	doi = {10.1088/1475-7516/2021/01/072},
	pages = {072--072},
	number = {1},
	journaltitle = {Journal of Cosmology and Astroparticle Physics},
	shortjournal = {J. Cosmol. Astropart. Phys.},
	author = {Giese, Felix and Konstandin, Thomas and Schmitz, Kai and de Vis, Jorinde van},
	date = {2021-01-29}
}

@article{lecture_notes,
	title = {Phase transitions in the early universe},
	issn = {2590-1990},
	url = {https://scipost.org/10.21468/SciPostPhysLectNotes.24},
	doi = {10.21468/SciPostPhysLectNotes.24},
	pages = {24},
	journaltitle = {{SciPost} Physics Lecture Notes},
	shortjournal = {{SciPost} Phys. Lect. Notes},
	author = {Hindmarsh, Mark and Lüben, Marvin and Lumma, Johannes and Pauly, Martin},
	date = {2021-02-15},
}

@article{gowling_lisa_2021,
	title = {Observational prospects for phase transitions at {LISA}: Fisher matrix analysis},
	volume = {2021},
	issn = {1475-7516},
	url = {https://iopscience.iop.org/article/10.1088/1475-7516/2021/10/039},
	doi = {10.1088/1475-7516/2021/10/039},
	shorttitle = {Observational prospects for phase transitions at {LISA}},
	pages = {039},
	number = {10},
	journaltitle = {Journal of Cosmology and Astroparticle Physics},
	shortjournal = {J. Cosmol. Astropart. Phys.},
	author = {Gowling, Chloe and Hindmarsh, Mark},
	date = {2021-10-01}
}

@article{hoche_towards_2021,
	title = {Towards an all-orders calculation of the electroweak bubble wall velocity},
	volume = {2021},
	copyright = {http://iopscience.iop.org/info/page/text-and-data-mining},
	issn = {1475-7516},
	url = {https://iopscience.iop.org/article/10.1088/1475-7516/2021/03/009},
	doi = {10.1088/1475-7516/2021/03/009},
	number = {03},
	journal = {Journal of Cosmology and Astroparticle Physics},
	author = {Höche, Stefan and Kozaczuk, Jonathan and Long, Andrew J. and Turner, Jessica and Wang, Yikun},
	month = mar,
	year = {2021},
	pages = {009},
}

@article{tenkanen_speed_2022,
	title = {Speed of sound in cosmological phase transitions and effect on gravitational waves},
	volume = {2022},
	issn = {1029-8479},
	url = {https://link.springer.com/10.1007/JHEP08(2022)302},
	doi = {10.1007/JHEP08(2022)302},
	pages = {302},
	number = {8},
	journaltitle = {Journal of High Energy Physics},
	shortjournal = {J. High Energ. Phys.},
	author = {Tenkanen, Tuomas V. I. and Van De Vis, Jorinde},
	date = {2022-08-30},
	langid = {english},
}

@article{sharma_shallow_2023,
	title = {Shallow relic gravitational wave spectrum with acoustic peak},
	volume = {2023},
	issn = {1475-7516},
	url = {https://iopscience.iop.org/article/10.1088/1475-7516/2023/12/042},
	doi = {10.1088/1475-7516/2023/12/042},
	number = {12},
	journal = {Journal of Cosmology and Astroparticle Physics},
	author = {Sharma, Ramkishor and Dahl, Jani and Brandenburg, Axel and Hindmarsh, Mark},
	month = dec,
	year = {2023},
	pages = {042}
}

@misc{pol_characterization_2023,
	title = {Characterization of the gravitational wave spectrum from sound waves within the sound shell model},
	copyright = {Creative Commons Attribution 4.0 International},
	url = {https://arxiv.org/abs/2308.12943},
	doi = {10.48550/ARXIV.2308.12943},
	publisher = {arXiv},
	author = {Pol, Alberto Roper and Procacci, Simona and Caprini, Chiara},
	year = {2023},
	note = {Version Number: 3},
}

@misc{giombi_cs_2024,
	title = {Signatures of the speed of sound on the gravitational wave power spectrum from sound waves},
	rights = {{arXiv}.org perpetual, non-exclusive license},
	url = {https://arxiv.org/abs/2409.01426},
	doi = {10.48550/ARXIV.2409.01426},
	publisher = {{arXiv}},
	author = {Giombi, Lorenzo and Dahl, Jani and Hindmarsh, Mark},
	date = {2024},
	note = {Version Number: 1},
}

@article{tian_gw_2024,
	title = {Gravitational waves from cosmological first-order phase transitions with precise hydrodynamics},
	rights = {{arXiv}.org perpetual, non-exclusive license},
	url = {https://arxiv.org/abs/2409.14505},
	doi = {10.48550/ARXIV.2409.14505},
	publisher = {{arXiv}},
	author = {Tian, Chi and Wang, Xiao and Balázs, Csaba},
	date = {2024},
	note = {Version Number: 1},
}

@misc{colpi_lisa_2024,
	title = {{LISA} {Definition} {Study} {Report}},
	copyright = {arXiv.org perpetual, non-exclusive license},
	url = {https://arxiv.org/abs/2402.07571},
	doi = {10.48550/ARXIV.2402.07571},
	publisher = {arXiv},
	author = {Colpi, Monica and Danzmann, Karsten and Hewitson, Martin and Holley-Bockelmann, Kelly and Jetzer, Philippe and Nelemans, Gijs and Petiteau, Antoine and Shoemaker, David and Sopuerta, Carlos and Stebbins, Robin and Tanvir, Nial and Ward, Henry and Weber, William Joseph and Thorpe, Ira and Daurskikh, Anna and Deep, Atul and Núñez, Ignacio Fernández and Marirrodriga, César García and Gehler, Martin and Halain, Jean-Philippe and Jennrich, Oliver and Lammers, Uwe and Larrañaga, Jonan and Lieser, Maike and Lützgendorf, Nora and Martens, Waldemar and Mondin, Linda and Niño, Ana Piris and Amaro-Seoane, Pau and Sedda, Manuel Arca and Auclair, Pierre and Babak, Stanislav and Baghi, Quentin and Baibhav, Vishal and Baker, Tessa and Bayle, Jean-Baptiste and Berry, Christopher and Berti, Emanuele and Boileau, Guillaume and Bonetti, Matteo and Brito, Richard and Buscicchio, Riccardo and Calcagni, Gianluca and Capelo, Pedro R. and Caprini, Chiara and Caputo, Andrea and Castelli, Eleonora and Chen, Hsin-Yu and Chen, Xian and Chua, Alvin and Davies, Gareth and Derdzinski, Andrea and Domcke, Valerie Fiona and Doneva, Daniela and Dvorkin, Irna and Ezquiaga, Jose María and Gair, Jonathan and Haiman, Zoltan and Harry, Ian and Hartwig, Olaf and Hees, Aurelien and Heffernan, Anna and Husa, Sascha and Izquierdo, David and Karnesis, Nikolaos and Klein, Antoine and Korol, Valeriya and Korsakova, Natalia and Kupfer, Thomas and Laghi, Danny and Lamberts, Astrid and Larson, Shane and Jeune, Maude Le and Lewicki, Marek and Littenberg, Tyson and Madge, Eric and Mangiagli, Alberto and Marsat, Sylvain and Vilchez, Ivan Martin and Maselli, Andrea and Mathews, Josh and van de Meent, Maarten and Muratore, Martina and Nardini, Germano and Pani, Paolo and Peloso, Marco and Pieroni, Mauro and Pound, Adam and Quelquejay-Leclere, Hippolyte and Ricciardone, Angelo and Rossi, Elena Maria and Sartirana, Andrea and Savalle, Etienne and Sberna, Laura and Sesana, Alberto and Shoemaker, Deirdre and Slutsky, Jacob and Sotiriou, Thomas and Speri, Lorenzo and Staab, Martin and Steer, Danièle and Tamanini, Nicola and Tasinato, Gianmassimo and Torrado, Jesus and Torres-Orjuela, Alejandro and Toubiana, Alexandre and Vallisneri, Michele and Vecchio, Alberto and Volonteri, Marta and Yagi, Kent and Zwick, Lorenz},
	year = {2024},
	note = {Version Number: 1},
}

@misc{ptplot,
	title = {{PTPlot}: a tool for exploring the gravitational wave power spectrum from first-order phase transitions},
	copyright = {Creative Commons Attribution 4.0 International},
	shorttitle = {{PTPlot}},
	url = {https://zenodo.org/doi/10.5281/zenodo.6949106},
	publisher = {Zenodo},
	author = {Weir, David James},
	collaborator = {Hooper, Deanna and Häkkinen, Jenni},
	month = aug,
	year = {2023},
	doi = {10.5281/ZENODO.6949106},
}

@misc{pttools,
    title = {PTtools},
    url = {https://github.com/CFT-HY/pttools},
    author = {Hindmarsh, Mark and Hopling, Chloe and Mäki, Mika and Giombi, Lorenzo},
    date = {2015-2025},
    langid = {english},
    doi = {10.5281/zenodo.15268219},
    note = {v0.10.0},
}

@misc{rasanen_gr_2022,
	title = {General relativity lecture notes},
	url = {https://www.mv.helsinki.fi/home/syrasane/gr/},
	author = {Räsänen, Syksy},
	date = {2022},
	langid = {english}
}

@misc{thesis_source,
    title = {The effect of sound speed on the gravitational wave spectrum of first order phase transitions in the early universe, source code},
    url = {https://github.com/AgenttiX/msc-thesis2},
    author = {Mika Mäki},
    date = {2021-2025},
    langid = {english}
}



\end{document}